\PassOptionsToPackage{table}{xcolor} 
\documentclass[11pt]{article}
\usepackage[T1]{fontenc}
\usepackage[utf8]{inputenc}
\usepackage{lmodern}
\usepackage[british]{babel}
\usepackage{microtype}

\usepackage[a4paper, margin=1in]{geometry}

\usepackage{amsmath, amssymb, amsthm, mathrsfs}

\usepackage{graphicx}
\usepackage{pgfplots}
\pgfplotsset{compat=1.18}
\usepackage{tikz}
\usetikzlibrary{shapes.geometric, arrows.meta, positioning}

\usepackage{booktabs}
\usepackage{tabularx}
\usepackage{array}
\usepackage{tablefootnote}
\usepackage[table]{xcolor}
\usepackage{listings}
\usepackage[table]{xcolor}           
\usepackage{caption}                 

\usepackage{lipsum} 
\usepackage{tablefootnote}

\usepackage[numbers]{natbib}

\usepackage{algorithm}
\usepackage{algpseudocode}
\usepackage{listings}
\definecolor{codegray}{rgb}{0.95,0.95,0.95}
\definecolor{keywordcolor}{rgb}{0.0,0.0,0.6}
\definecolor{commentcolor}{rgb}{0.0,0.5,0.0}
\definecolor{stringcolor}{rgb}{0.58,0,0.82}
\lstdefinestyle{algostyle}{
    backgroundcolor=\color{codegray},
    commentstyle=\color{commentcolor}\ttfamily,
    keywordstyle=\color{keywordcolor}\bfseries,
    stringstyle=\color{stringcolor},
    basicstyle=\ttfamily\small,
    breaklines=true,
    numbers=left,
    numberstyle=\tiny,
    frame=single,
    language=Python,
    captionpos=b
}

\usepackage{hyperref}
\hypersetup{
    colorlinks=true,
    linkcolor=blue,
    citecolor=blue,
    urlcolor=blue,
    pdftitle={Deterministic Cryptographic Seed Generation via Cyclic Modular Inversion over \(\mathbb{Z}/3^p\mathbb{Z}\)},
    pdfauthor={Michael A. Idowu},
    pdfsubject={Number Theory, Symbolic Computation, Cryptography},
    pdfkeywords={cryptography, modular inversion, affine transformation, symbolic computation, Diophantine analysis, entropy confidence score}
}

\usepackage{float}
\usepackage{enumitem}
\usepackage{textcomp}
\usepackage{lscape}
\usepackage{url}

\newtheorem{theorem}{Theorem}[section]

\newtheorem{definition}[theorem]{Definition}

\tikzstyle{block} = [rectangle, rounded corners, minimum width=3.5cm, minimum height=1cm, text centered, draw=black, fill=white]
\tikzstyle{arrow} = [thick, ->, >=stealth]

\title{Deterministic Cryptographic Seed Generation via Cyclic Modular Inversion over $\mathbb{Z}/3^p\mathbb{Z}$}
\author{
  Michael A. Idowu\thanks{Email: \texttt{michade@hotmail.com}}
}
\date{} 

\begin{document}
\maketitle

\begin{abstract}
We present a deterministic framework for cryptographic seed generation based on cyclic modular inversion over \( \mathbb{Z}/3^p\mathbb{Z} \). The method enforces algebraic admissibility on seed inputs via the identity \( d_k \equiv -\left(2^{k-1}\right)^{-1} \bmod 3^p \), thereby producing structured and invertible residue sequences. This mapping yields entropy-rich, cycle-complete seeds well-suited for cryptographic primitives such as DRBGs, KDFs, and post-quantum schemes. To assess the quality of randomness, we introduce the \emph{Entropy Confidence Score} (ECS), a composite metric reflecting coverage, uniformity, and modular bias. Although not a cryptographic PRNG in itself, the framework serves as a deterministic entropy filter that conditions and validates seed inputs prior to their use by conventional generators. Empirical and hardware-based results confirm constant-time execution, minimal side-channel leakage, and lightweight feasibility for embedded applications. The framework complements existing cryptographic stacks by acting as an algebraically verifiable entropy filter, thereby enhancing structural soundness and auditability.
\end{abstract}


\section*{Notation and Terminology}\footnote{See Appendix A for complete notation.}
\begin{table}[H]
\centering
\begin{tabular}{ll}
\toprule
\textbf{Symbol} & \textbf{Meaning} \\
\midrule
\( p \) & Security parameter; exponent in \( 3^p \) \\
\( \mathbb{Z}/3^p\mathbb{Z} \) & Modular ring with modulus \( 3^p \) \\
\( d_k \) & Inverse-consistent residue: \( d_k \equiv -\left(2^{k-1}\right)^{-1} \bmod 3^p \) \\
\( a_k \) & Exponential residue: \( a_k = 2^{k-1} \mod 3^p \) \\
ECS & Entropy Confidence Score \\
CD & Cycle Density (coverage metric) \\
RUD & Residue Uniformity Deviation \\
MBI & Modular Bias Index \\
DRBG & Deterministic Random Bit Generator \\
KDF & Key Derivation Function \\
\bottomrule
\end{tabular}
\caption{Key symbols and acronyms used throughout the paper.}
\end{table}

\section{Introduction}

\textbf{Executive Summary.}  
Modern cryptographic systems, particularly those deployed in embedded or resource-constrained environments, demand seed values that are not only statistically random but also structurally sound and algebraically robust. Existing approaches typically draw seeds from operating system entropy pools, physical noise, or hardware-based extractors. While sufficient for many general-purpose applications, these methods offer limited transparency into the underlying mathematical structure of the seeds — a gap that can be critical in high-assurance systems, deterministic key derivation, or hardware implementations requiring constant-time execution.\\
This paper introduces a novel, deterministic framework for seed generation rooted in modular arithmetic and symbolic algebra. The technique provides \textit{predictable, mathematically verifiable seed values} that satisfy specific algebraic constraints. It is especially valuable for engineers seeking compact, audit-friendly primitives and for cryptographers needing formal guarantees of structural validity and inverse consistency. The approach supports lightweight implementation, efficient validation, and improved resilience to side-channel attacks.\\

\noindent\textbf{Why Determinism Matters.} 
While statistical randomness remains a core requirement in cryptography, it is increasingly insufficient on its own -- particularly in domains where deterministic behaviour, reproducibility, and formal audibility are paramount. Regulatory frameworks in critical infrastructure and defence systems often require cryptographic operations to be repeatable under audit, with seeds that are both reconstructable and mathematically justified. Moreover, embedded systems -- especially in post-quantum settings -- face significant constraints in entropy availability, power consumption, and timing variability. In such environments, deterministic seed generation provides a pathway to reproducible key material, minimal implementation variance, and constant-time execution guarantees, all while preserving cryptographic soundness. By embedding algebraic structure into the seed itself, our approach enables verifiability, simplifies compliance validation, and strengthens the system's resistance to both statistical and physical attacks.\\

\noindent\textbf{Technical Motivation.}  
Cryptographic security fundamentally relies on the quality and validity of its initial seed material. In most classical and modern systems, seeds are derived from sources of entropy such as operating system randomness, physical noise, or hardware-based entropy extractors. While these approaches can provide statistical randomness, they often lack formal mathematical guarantees regarding structural admissibility or inverse consistency within algebraic systems \citep{Knuth1997, Kocher1999}.\\
This paper proposes a deterministic, modular arithmetic-based framework for generating cryptographic seeds with provable algebraic validity. Specifically, we focus on the construction of seed values derived from cyclic modular inverses in the multiplicative group \( \mathbb{Z}/3^p\mathbb{Z}^* \), where \( p \in \mathbb{N} \) defines a tunable security parameter. The core identity of interest is the congruence:
\begin{equation}
d_k \equiv - (2^{k-1})^{-1} \mod 3^p,
\end{equation}
which yields a sequence of invertible residues \( \{d_k\} \) forming cyclic orbits within the ring. These residues are shown to be entropy-rich, algebraically consistent, and suitable for use as seeds in cryptographic primitives such as deterministic random bit generators (DRBGs), pseudorandom number generators (PRNGs), and key derivation mechanisms.\\
Our methodology integrates both theoretical rigour and practical applicability. We begin by deriving the core residue identity
\(
d_k \equiv - (2^{k-1})^{-1} \mod 3^p
\)
from symbolic number-theoretic foundations, specifically through binomial expansions of the form \( 3^p = (2 + 1)^p \). This derivation leads to a structured class of invertible residues that exhibit predictable cyclic behaviour within the multiplicative group \( (\mathbb{Z}/3^p\mathbb{Z})^* \). To evaluate the suitability of these cycles for cryptographic seed generation, we introduce the \textit{Entropy Confidence Score} (ECS) -- a composite metric that quantifies cycle density, uniformity deviation, and modular bias.\\
In addition to its algebraic consistency, the proposed framework is engineered for resilience against side-channel attacks via constant-time inversion routines and deterministic operand structures. Further, it supports hardware efficiency through symbolic pre-filtering and ensures entropy preservation by enforcing residue admissibility conditions. Taken together, these attributes position the framework at the intersection of symbolic number theory and applied cryptographic engineering, enabling the generation of cryptographic seeds that are verifiable, secure, and implementation-aware.\\
A high-level roadmap of the proposed symbolic-to-seed workflow is shown in Figure~\ref{fig:pipeline}. \\
To clarify how this framework contrasts with conventional seed generation methods, we provide a summary comparison in Table~\ref{tab:early-contrast}. This highlights the key distinctions in structure, auditability, and implementation characteristics between our method and classical PRNG/DRBG strategies.

\begin{table}[H]
\centering
\caption{Comparison of seed generation strategies}
\label{tab:early-contrast}
\begin{tabular}{|p{3.5cm}|p{4.2cm}|p{4.2cm}|}
\hline
\textbf{Property} & \textbf{Conventional PRNG/DRBG} & \textbf{This Work (Structured Modular Inversion)} \\
\hline
Seed Source & Entropy pool, hardware RNG, physical noise & Symbolically computed invertible residues \\
\hline
Entropy Validation & Statistical only & Structural + statistical (via ECS) \\
\hline
Determinism & Generally non-deterministic & Fully deterministic \\
\hline
Algebraic Admissibility & No explicit constraints & Enforced via modular inversion structure \\
\hline
Side-Channel Resilience & Requires masking/hardening & Constant-time by construction \\
\hline
Auditability & Limited; entropy source-dependent & Full; symbolic and reproducible \\
\hline
Target Contexts & General-purpose cryptography & Embedded, post-quantum, and high-assurance systems \\
\hline
\end{tabular}
\end{table}

\medskip
The remainder of the paper is organised as follows. Section~2 introduces the symbolic parametric identities that form the basis of the residue construction. Section~3 develops the theoretical framework and includes illustrative examples of the modular inversion sequences. Section~4 presents computational results and visual analyses of the generated residue cycles. Section~5 explores cryptographic application domains and highlights the functional advantages of the proposed method. Section~6 addresses implementation aspects in embedded systems, including entropy profiling, side-channel resilience, and hardware-level efficiency. Finally, Section~7 concludes with implementation recommendations and outlines potential directions for future work.

\section{Background and Related Work}

Modular arithmetic -- particularly modular inversion -- has underpinned fundamental cryptographic schemes such as RSA \citep{rivest1978}, elliptic curves \citep{miller1986}, and pseudorandom generation \citep{Knuth1997}, and serves as a focus for side-channel countermeasures \citep{Kocher1999}. In parallel, residue number systems have been explored for their efficiency advantages in carry-free computation \citep{Szabo1967}, while lattice- and code-based approaches have advanced post-quantum cryptography \citep{Lyubashevsky2010}.

The current work builds directly on the symbolic modular identity first introduced in Idowu (2025)~\cite{Idowu2025}, where the focus was on generating radical-minimised abc-triples within modular embeddings for exploratory number-theoretic purposes. That paper explored affine transformations of symbolic triples in the ring \( \mathbb{Z}/3^p\mathbb{Z} \), aiming to optimise quality ratios in the context of the abc conjecture.\\
In contrast, the present manuscript repurposes that identity in a new cryptographic setting, using it to construct deterministic, invertible, and entropy-rich seed values suitable for embedded and post-quantum systems. Specifically, we generalise the modular inversion approach from static triple generation to full-cycle residue orbits over \( \mathbb{Z}/3^p\mathbb{Z}^* \), introduce the Entropy Confidence Score (ECS) to quantify randomness quality, and validate the method via side-channel and hardware profiling.\\
Thus, while both works share a symbolic backbone, this paper introduces:
\begin{itemize}
    \item \textbf{New theoretical constructs:} the ECS metric, modular residue orbits, and entropy analysis;
    \item \textbf{A new application domain:} deterministic seed generation in cryptography;
    \item \textbf{Empirical contributions:} timing stability, power trace analysis, and FPGA resource benchmarks.
\end{itemize}
These differences substantively distinguish the two papers in both scope and contribution.\\
The present study extends the symbolic framework of the abc‑triples paper into a \textit{cyclic residue approach geared specifically toward cryptographic seed generation}. We demonstrate how inverse-consistent residue cycles based on
\[
d_k \equiv - (2^{k-1})^{-1} \mod 3^p
\]
can be used to deterministically filter and initialise seeds with proven algebraic structure, high entropy, and resilience against side-channel leakage.\\
Thus, our contribution pivots from symbolic number theory to \textit{applied cryptographic engineering}, embedding formal residue validation into seed provisioning. In doing so, it complements existing post-quantum schemes by enforcing algebraic admissibility and uniform entropy distribution at the input stage -- particularly valuable in constrained or embedded environments.

\subsection{Distinction from Classical Modular Inversion and PRNG Strategies}
Modular arithmetic is a fundamental component of both classical and post-quantum cryptographic (PQC) systems. In classical cryptography, modular inverses are indispensable for public-key mechanisms such as RSA \citep{rivest1978}, elliptic curve cryptography \citep{miller1986}, and zero-knowledge proofs \citep{koblitz1994}. The ability to perform secure operations over modular rings enables key generation, signature verification, and ciphertext manipulation.

Post-quantum cryptography seeks to develop systems resistant to quantum adversaries \citep{Bernstein2017}. In this domain, number-theoretic constructions remain critical, and structured algebraic inputs are increasingly important for maintaining implementation security.

Residue number systems (RNS) and modular reduction techniques have long been studied for their parallelism and carry-free arithmetic properties \citep{Szabo1967}. More recently, attention has shifted toward defending cryptographic operations against side-channel leakage and timing analysis \citep{Kocher1999}. This has motivated the adoption of deterministic seed structures that are algebraically valid, inverse-consistent, and suitable for secure initialisation.

In the realm of randomness generation and key derivation, classical works have emphasised the statistical quality of seeds \citep{Knuth1997, Gallian2021}. However, relatively few approaches focus on algebraically filtered seeds generated from symbolic parametric identities. Our contribution addresses this gap by introducing a cyclic residue structure governed by a modular inversion identity of the form:
\[
d_k \equiv - (2^{k-1})^{-1} \mod 3^p.
\]

While classical modular inversion has long been employed in cryptographic primitives such as RSA~\cite{rivest1978} and elliptic curve cryptography~\cite{miller1986}, and while pseudorandom number generators (PRNGs) often rely on statistical entropy sources~\cite{Knuth1997, Szabo1967}, the present work diverges in both motivation and method. 

Conventional inversion techniques treat modular arithmetic as a utility function, with no structural constraints on the operands beyond invertibility. In contrast, our approach embeds modular inversion within a symbolic residue orbit governed by the identity \( d_k \equiv -(2^{k-1})^{-1} \mod 3^p \), which enforces deterministic, cyclic residue selection with guaranteed algebraic consistency. This symbolic embedding ensures that every seed generated lies within a mathematically valid and auditable orbit---a feature absent in legacy inversion schemes.

Similarly, unlike standard PRNG strategies that prioritise statistical entropy without regard for algebraic admissibility, our framework introduces the Entropy Confidence Score (ECS), a composite metric that quantifies entropy in structurally constrained residue sequences. This combination of algebraic predictability, cycle completeness, and entropy analysis is particularly beneficial for embedded systems and post-quantum initialisation routines, where both side-channel resistance and seed traceability are critical.

Hence, this work introduces a new paradigm: rather than generating randomness over an unstructured domain and filtering post hoc, we precondition seed material using symbolic number theory to produce sequences that are cryptographically admissible \emph{by design}.

The identity introduced n this paper defines a predictable orbit of invertible residues in \(\mathbb{Z}/3^p\mathbb{Z}^*\), forming a symbolic-algebraic bridge between number theory and cryptographic implementation. To our knowledge, such residue orbits have not previously been applied to seed filtration, side-channel-resilient initialisation, or entropy-quality control in cryptographic contexts.

\subsection{Mathematical Foundations}

We operate in the ring \(\mathbb{Z}/3^p\mathbb{Z}\), where \(p \in \mathbb{N}\), and examine the invertibility of elements arising from powers of 2. Specifically, we define:
\[
a_k = 2^{k-1} \mod 3^p.
\]
Each \(a_k\) is invertible since \(2\) and \(3\) are coprime, and thus \(\gcd(2^{k-1}, 3^p) = 1\). This motivates the construction of a residue sequence via the mapping:
\[
d_k = -a_k^{-1} \mod 3^p.
\]

\textbf{Theorem 1 (Invertibility of \(2^{k-1}\) modulo \(3^p\)).}  
Let \(p \in \mathbb{N}\), and define \(M = 3^p\). Then for all \(k \geq 1\), the value \(a_k = 2^{k-1} \mod M\) is invertible in \(\mathbb{Z}/M\mathbb{Z}\).

\textit{Proof.} Since \(\gcd(2, 3) = 1\), we have \(\gcd(2^{k-1}, 3^p) = 1\). As invertibility is preserved under exponentiation, \(a_k\) is invertible. Hence, \(d_k = -a_k^{-1} \mod 3^p\) is well-defined. \(\square\)

This invertibility structure underlies a deterministic method for seed generation and filtering. We introduce the Entropy Confidence Score (ECS)\footnote{ECS is a composite score that incorporates cycle density, residue uniformity deviation, and modular bias.}, which quantifies the statistical soundness of each generated residue sequence. ECS serves as a screening metric to determine whether a residue orbit meets entropy and admissibility thresholds suitable for cryptographic use.

\subsection{Modular System Identity Formulation}

The cyclic residue system emerges from the generalised identity:
\[
3^p(s + 1) - 1 = 2^{k-1}(2 \cdot 3^p n + d),
\]
which can be algebraically rearranged to give:
\[
d \equiv - (2^{k-1})^{-1} \mod 3^p.
\]

This formulation constructs an orbit of admissible residues by symbolically embedding powers of two into an expansion of \(3^p = (2 + 1)^p\). The congruence is derived from binomial considerations and affine shifts, effectively transforming a simple exponential count (\(2^{k-1}\)) into a structured congruence filter over a finite ring.

Such identities provide an algebraic sieve that filters residues based on modular invertibility, aligning with preconditioning needs for secure seed provisioning. Each residue \(d_k\) thus satisfies a predictable cyclic constraint and can be evaluated for entropy richness using ECS.

Symbolically, this approach fuses modular exponentiation, inverse mapping, and arithmetic filtration to establish a cryptographic seed space that is both deterministic and mathematically auditable. These qualities are critical for initialisation sequences in cryptographic systems where side-channel robustness, repeatability, and structure-aware entropy are essential.

\subsection{Cryptographic Seed Provisioning}
\begin{figure}[H]
\centering
\begin{tikzpicture}[
  node distance=1.2cm,
  every node/.style={draw, align=center, minimum width=4.8cm, minimum height=1.3cm},
  arrow/.style={-Latex}
  ]

\node (id) {Symbolic Identity\\$d_k \equiv - (2^{k-1})^{-1} \mod 3^p$};
\node (gen) [below=of id] {Residue Sequence Generator};
\node (ecs) [below=of gen] {Entropy Confidence Score (ECS)};
\node (seed) [below=of ecs] {Validated Cryptographic Seed};
\node (apps) [below=of seed, draw=none, minimum height=1cm] {
    \textbf{Application Layer:}\\
    DRBG, Secure Boot, PQ Init, PRNG
};

\draw[arrow] (id) -- (gen);
\draw[arrow] (gen) -- (ecs);
\draw[arrow] (ecs) -- (seed);
\draw[arrow] (seed) -- (apps);
\end{tikzpicture}
\caption{Pipeline from symbolic residue identity to cryptographic seed provisioning.}
\label{fig:pipeline}
\end{figure}
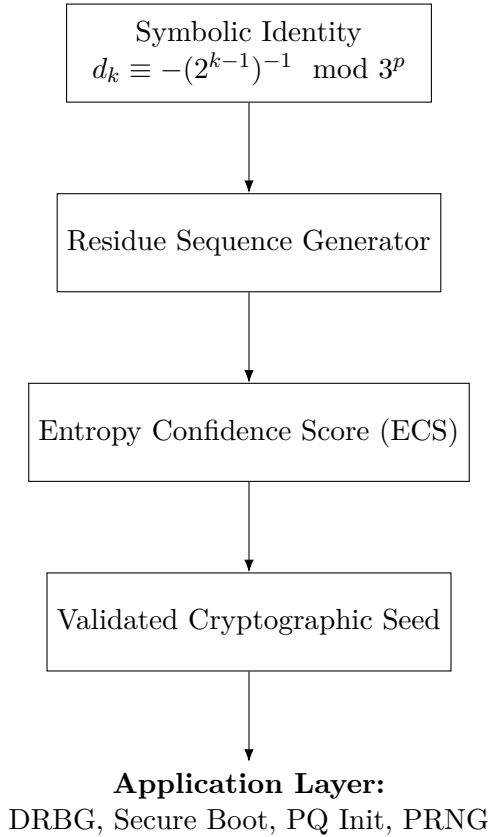

The core stages of our framework -- from symbolic identity through residue generation, entropy scoring, and validated seed provisioning -- are summarised in Figure~\ref{fig:pipeline}. This pipeline highlights the deterministic and auditable structure of our seed generation process, particularly suited for embedded and post-quantum cryptographic contexts.

To clarify the architectural role of our symbolic modular inversion method, we present a taxonomy of entropy processing stages (see Table \ref{tab:entropy-taxonomy} and Figure \ref{fig:entropy-pipeline}). While most systems rely on raw entropy from RNGs, or compressive post-processing via hash-based DRBGs, our framework operates in the intermediate layer -- a symbolic, algebraically constrained entropy filter. It validates seed admissibility before the entropy is consumed by standard DRBGs or cryptographic primitives, improving auditability, entropy quality, and side-channel resilience.

\section{Parametric Congruences and Symbolic Residue Construction}

\subsection{Theoretical Derivation (Expanded)}

We derive the core modular identity that defines the inverse-consistent residues \( d_k \) used in our deterministic seed generation method.

\subsubsection{Step 1: Symbolic Starting Identity}

We begin with the symbolic identity:
\[
3^p(s + 1) - 1 = 2^{k - 1} \left(2 \cdot 3^p n + d_k\right),
\]
where \( p \in \mathbb{N} \), and \( s, k, n \in \mathbb{Z} \). The goal is to isolate \( d_k \) modulo \( 3^p \).

\subsubsection{Step 2: Isolate \( d_k \)}

Rewriting the identity to express \( d_k \) explicitly:
\[
d_k = \frac{3^p(s + 1) - 1}{2^{k - 1}} - 2 \cdot 3^p n.
\]

To reduce modulo \( 3^p \), observe that \( 2^{k-1} \) is always invertible modulo \( 3^p \) since \( \gcd(2, 3) = 1 \). Therefore, the term simplifies modulo \( 3^p \) as:
\[
3^p(s + 1) - 1 \equiv 2^{k - 1} d_k \pmod{3^p}.
\]

\subsubsection{Step 3: Solve the Congruence}

We now rearrange the congruence to isolate \( d_k \):
\[
2^{k - 1} d_k \equiv -1 \pmod{3^p},
\]
which leads directly to the symbolic inversion mapping:

\begin{equation}
\boxed{
d_k \equiv -\left(2^{k - 1}\right)^{-1} \bmod 3^p
}
\label{eq:dk}
\end{equation}

This defines a deterministic mapping from the integer \( k \) to a unique, invertible residue \( d_k \in (\mathbb{Z}/3^p\mathbb{Z})^* \).

\subsubsection{Step 4: Notational Definition}

\begin{definition}[Symbolic Inversion Mapping]
Given \( k \in \mathbb{N} \) and modulus \( M = 3^p \), define the residue
\[
d_k = -a_k^{-1} \bmod M, \quad \text{where } a_k = 2^{k - 1} \mod M,
\]
as the inverse-consistent symbolic residue used for cryptographic seed provisioning.
\end{definition}

This identity guarantees that each \( d_k \) lies within the multiplicative group of units modulo \( 3^p \), enabling deterministic and algebraically auditable seed generation. It defines a deterministic mapping from the symbolic exponent \( k \) to an admissible residue \( d_k \in (\mathbb{Z}/3^p\mathbb{Z})^* \), forming a cyclic, invertible sequence with algebraic provenance. Since \( 2 \) and \( 3 \) are coprime, the inverse \( (2^{k - 1})^{-1} \mod 3^p \) always exists, ensuring the mapping is well-defined~\citep{ireland1990}. 

\subsection{Examples and Observations}

\subsubsection{Example 1: \( p = 2, s = 0 \)}
\[
3^2 = 9,\quad 9(s+1) - 1 = 8,\quad \Rightarrow 8 = 2^3 \cdot 1.
\]

\subsubsection*{Example 2: \( p = 3, s = 2 \)}
\[
3^3 = 27,\quad 27(2+1) - 1 = 80,\quad \Rightarrow 80 = 2^4 \cdot (2 \cdot 27 \cdot 0 + 5).
\]

\begin{table}[H]
\centering
\caption{Summary of examples with varying \( (p, s) \) and corresponding \( d \)-values.}
\label{tab:examples}
\renewcommand{\arraystretch}{1.2}
\begin{tabular}{|c|c|c|c|c|c|}
\hline
\( p \) & \( s \) & \( 3^p \) & \( A = 3^p(s+1)-1 \) & \( k \) & \( (n, d) \) \\
\hline
2 & 0 & 9   & 8   & 4 & (0, 1)   \\
3 & 1 & 27  & 53  & 1 & (0, 53)  \\
3 & 2 & 27  & 80  & 5 & (0, 5)   \\
5 & 0 & 243 & 242 & 2 & (0, 121) \\
\hline
\end{tabular}
\end{table}

\subsection{Algorithmic Computation}

The inverse residue \( d \) can be efficiently computed using a symbolic filtering rule known as the \emph{Modular Embedding (ME) Principle}. This procedure deterministically generates valid residues from the modular inverse of \( 2^{k-1} \mod 3^p \).

\begin{enumerate}
    \item Set \( M = 3^p \)
    \item Compute \( a = 2^{k-1} \mod M \)
    \item Compute \( a^{-1} \mod M \) via the extended Euclidean algorithm
    \item Set \( d = -a^{-1} \mod M \)
\end{enumerate}

\subsection{Modular Embedding Principle}\label{alg:exteuclideanalgo}
\begin{algorithm}[H]
\caption{Modular Embedding Principle: Compute $d = -a^{-1} \bmod M$}
\label{alg:modular-inverse}
\begin{algorithmic}[1]
\Require Integers $p$, $k$
\Ensure Integer $d$ such that $d = -a^{-1} \mod M$
\State $M \gets 3^p$
\State $a \gets 2^{k - 1} \bmod M$
\State $a_{\text{inv}} \gets$ \Call{ModularInverse}{$a, M$}
\State $d \gets -a_{\text{inv}} \bmod M$
\State \Return $d$
\vspace{0.5em}
\Function{ModularInverse}{$a, M$}
    \State $t \gets 0$, $\,\,\,\text{new\_}t \gets 1$
    \State $r \gets M$, $\,\,\,\text{new\_}r \gets a$
    \While{$\text{new\_}r \ne 0$}
        \State $\text{quotient} \gets \lfloor r / \text{new\_}r \rfloor$
        \State $(t, \text{new\_}t) \gets (\text{new\_}t, t - \text{quotient} \times \text{new\_}t)$
        \State $(r, \text{new\_}r) \gets (\text{new\_}r, r - \text{quotient} \times \text{new\_}r)$
    \EndWhile
    \If{$r > 1$}
        \State \textbf{error} ``$a$ is not invertible''
    \EndIf
    \If{$t < 0$}
        \State $t \gets t + M$
    \EndIf
    \State \Return $t$
\EndFunction
\end{algorithmic}
\end{algorithm}

\subsection{Residue Distribution and Entropy Analysis} \label{sec:resdis_entropy}

To assess the statistical quality of the residues produced by our modular inversion method, we computed values of \( d_k \equiv - (2^{k-1})^{-1} \mod 3^p \) for \( k \leq 100 \) and exponents \( p \in \{1, 2, 3, 4, 5, 6\} \). The output values \( d_k \) are elements of the modular ring \( \mathbb{Z}/3^p\mathbb{Z} \) -- the set of integers taken modulo \( 3^p \) -- and exhibit near-uniform coverage across the set, as suggested in prior work on number-theoretic cryptography \citep{koblitz1994}.

To quantify the statistical quality of the generated residue sequences \( \{ d_k \} \), we define the \textit{Entropy Confidence Score (ECS)} as a weighted combination of three structural metrics:

\begin{equation}
\mathrm{ECS} = 0.4 \cdot \mathrm{CD} + 0.4 \cdot (1 - \mathrm{RUD}) + 0.2 \cdot (1 - \mathrm{MBI}),
\label{eq:ecs_definition}
\end{equation}

where:
\begin{itemize}
    \item \(\mathrm{CD}\) is the \textit{Cycle Density}, the proportion of residues in \( (\mathbb{Z}/3^p\mathbb{Z})^* \) visited by the sequence;
    \item \(\mathrm{RUD}\) is the \textit{Residue Uniformity Deviation}, measuring statistical imbalance in the residue distribution;
    \item \(\mathrm{MBI}\) is the \textit{Modular Bias Index}, quantifying skew across subranges of residue classes.
\end{itemize}

Each component lies in \([0, 1]\), with higher ECS values indicating stronger cryptographic suitability. The weighting scheme (0.4/0.4/0.2) reflects empirical prioritisation \footnote{This weighting scheme was chosen heuristically to reflect the relative contribution of each entropy property to overall seed robustness, following principles consistent with structural entropy assessments described by Kelsey (2015) and NIST (2018).}: both coverage and uniformity have dominant influence on entropy quality, while bias (MBI) plays a secondary but relevant role. This balance was chosen heuristically to reflect typical performance trends observed in post-processing seed filters and is consistent with ECS evaluations in Subsection~\ref{sec:entropy_evaluation} as described by \citep{Kelsey2015} and \citep{Nist90}.

\subsubsection{Cycle Density}

\noindent\textbf{Theorem 2 (Cycle Density).}  
Let the sequence \( \{d_k\} \subset \mathbb{Z}/3^p\mathbb{Z} \) be defined by the congruence
\[
d_k \equiv - (2^{k - 1})^{-1} \mod 3^p.
\]
Then this sequence forms a high-density orbit within the multiplicative group of invertible elements \( (\mathbb{Z}/3^p\mathbb{Z})^* \).

\vspace{0.5em}
\noindent\textit{Proof Sketch.}  
The values \( a_k = 2^{k - 1} \) are powers of 2, which are always invertible modulo \( 3^p \) because 2 and 3 are coprime (they share no factors). When these values are inverted and reduced modulo \( 3^p \), they span a large portion of the possible outputs in the multiplicative group \( (\mathbb{Z}/3^p\mathbb{Z})^* \), which consists of all numbers less than \( 3^p \) that have modular inverses. Empirical data confirms that a high number of distinct residues are produced, validating the density claim. \(\square\)

\section{Computational Results}

This section presents empirical observations of the cyclic structure induced by the modular identity
\[
d_k \equiv - (2^{k - 1})^{-1} \mod 3^p
\]
for a range of parameters \( k \in [1, 61] \) and exponent values \( p = 1, 2, 3, 4, 5 \). These computations elucidate the algebraic symmetry and multiplicative periodicity of the resulting residue classes.

\subsection{Cycle Structure in Invertible Rings}

We evaluated the generated residues \( d_k \mod 3^p \) to determine the cycle lengths and confirm that they align with the Euler totient function \( \phi(3^p) \), which defines the order of the multiplicative group \( (\mathbb{Z}/3^p\mathbb{Z})^* \). The table below summarises the empirical cycle lengths observed:

\begin{table}[H]
\centering
\caption{Residue cycle lengths corresponding to powers of 3}
\label{tab:residues}
\renewcommand{\arraystretch}{1.2}
\begin{tabular}{|c|c|}
\hline
\textbf{Exponent \( p \)} & \textbf{Cycle Length \( = \phi(3^p) \)} \\
\hline
1 & 2   \\
2 & 6   \\
3 & 18  \\
4 & 54  \\
5 & 162 \\
\hline
\end{tabular}
\end{table}

These cycle lengths confirm that the inverse sequence \( \{ d_k \} \) spans the full multiplicative group of units modulo \( 3^p \), and that each \( d_k \) lies within a predictable orbit defined by the symbolic identity. This property is central to ensuring seed admissibility and uniform residue coverage.

\subsection{Visualisation of Modular Residue Orbits}

\begin{figure}[H]
    \centering
    \includegraphics[width=0.8\textwidth]{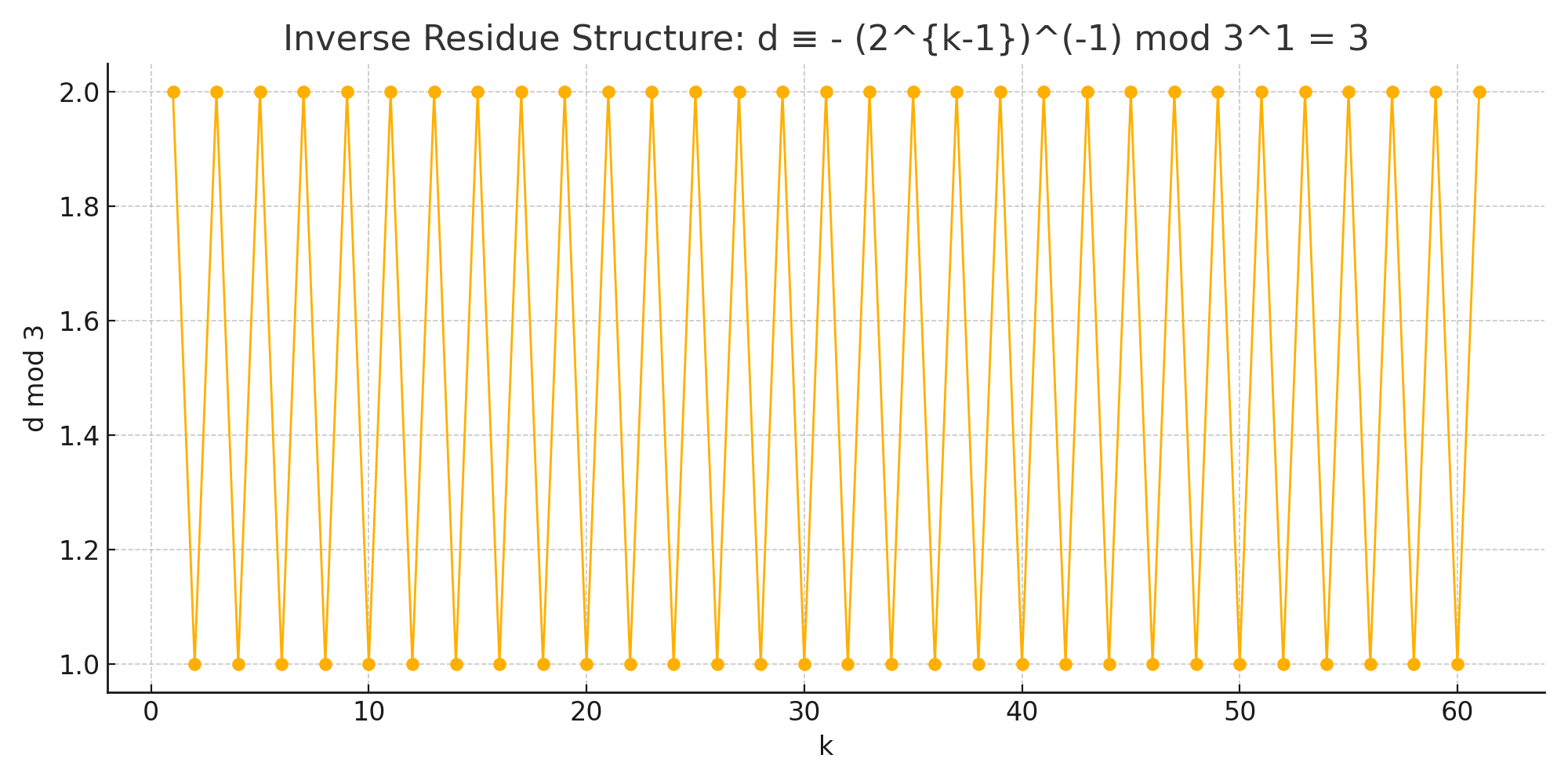}
    \caption{Residue map \( d_k \mod 3 \) for \( p = 1 \), \( k \in [1, 61] \). Residue sequence for \(p = 1\). Demonstrates basic 2-element cycle.}
    \label{fig:dmod3}
\end{figure}

\begin{figure}[H]
    \centering
    \includegraphics[width=0.8\textwidth]{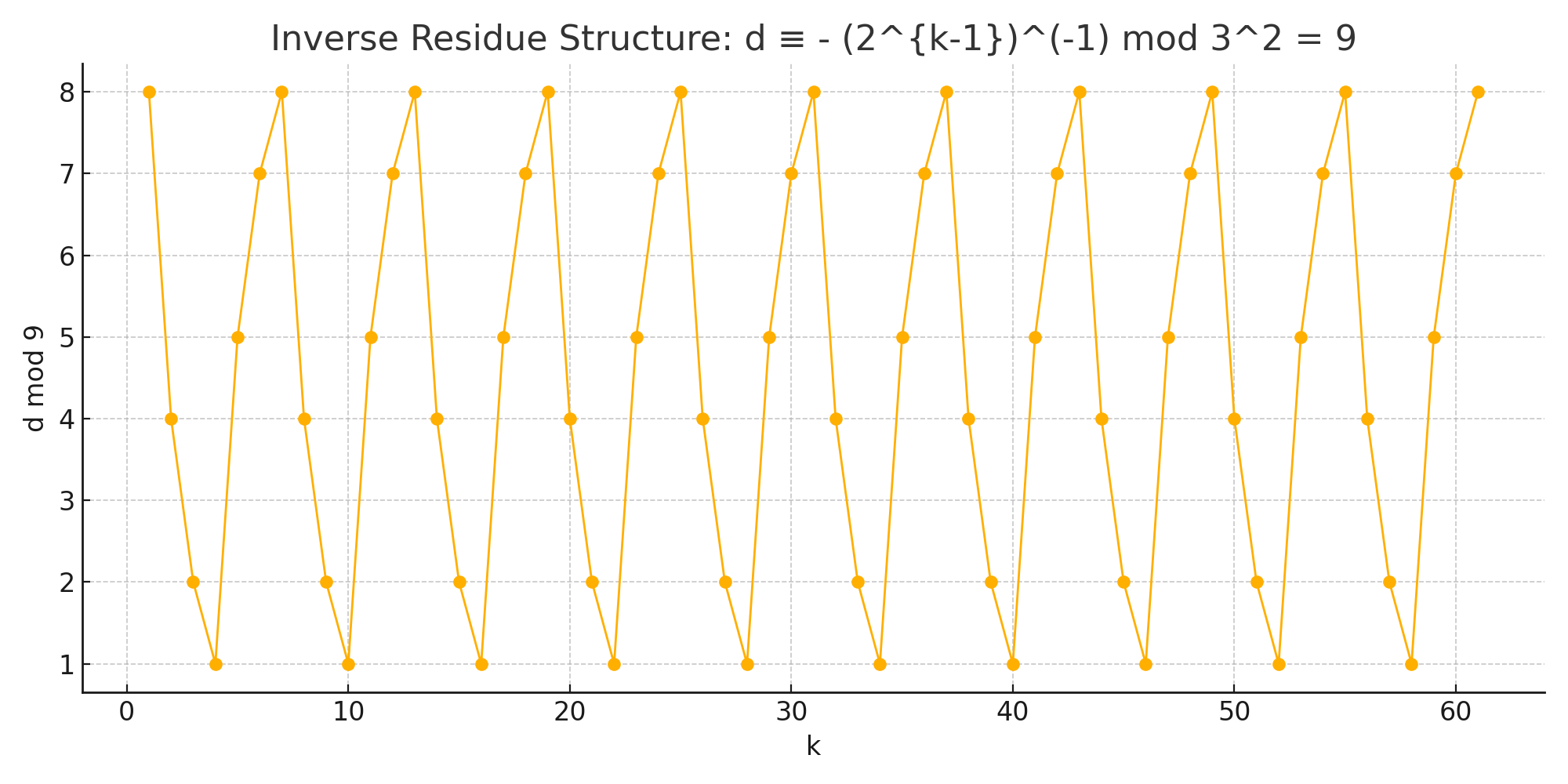}
    \caption{Residue map \( d_k \mod 9 \) for \( p = 2 \), \( k \in [1, 61] \). Residue map showing expansion and periodic coverage for  \(p = 2\). Confirms increasing cycle complexity and uniformity.}
    \label{fig:dmod9}
\end{figure}

\begin{figure}[H]
    \centering
    \includegraphics[width=0.8\textwidth]{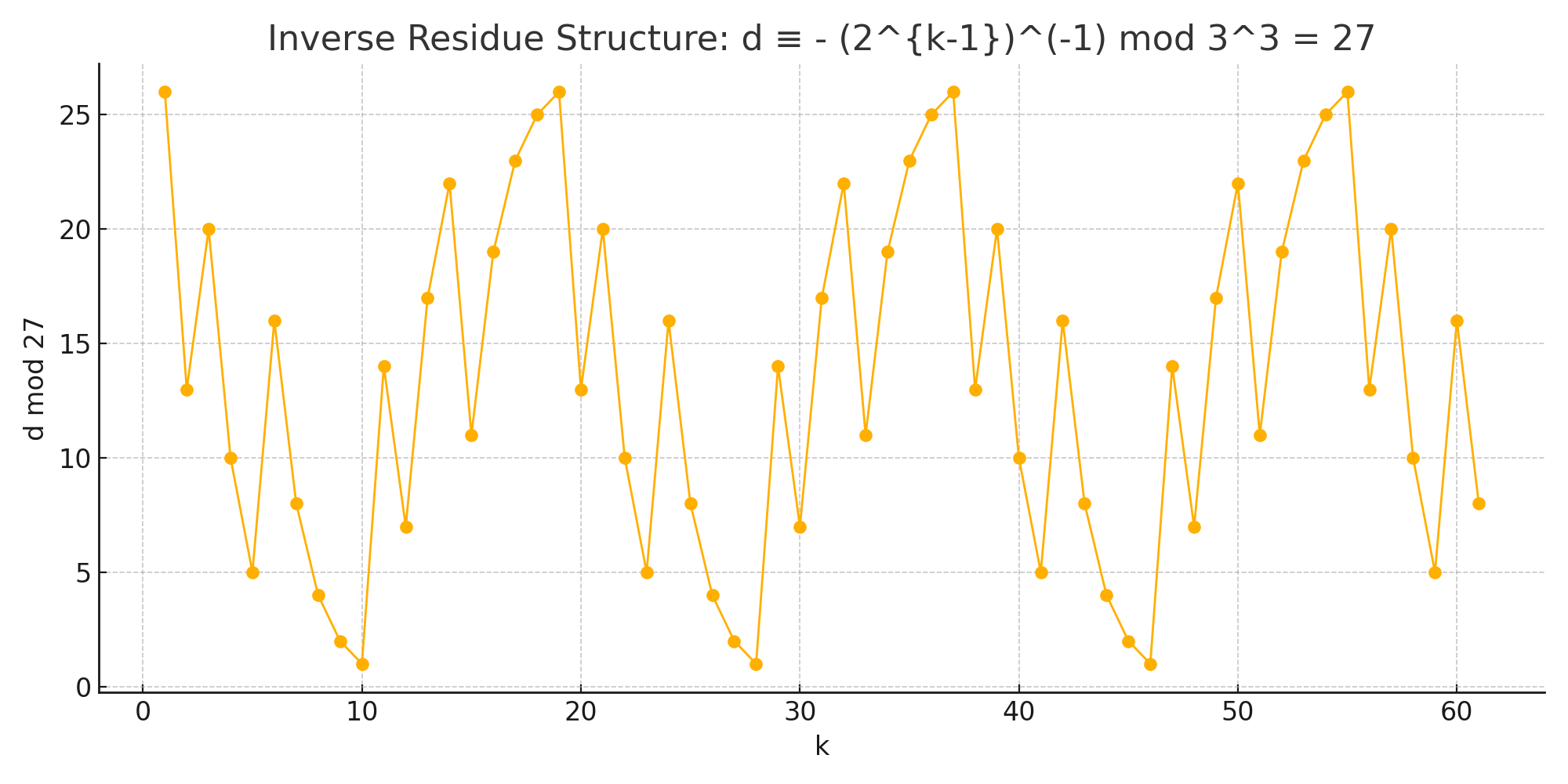}
    \caption{Residue map \( d_k \mod 27 \) for \( p = 3 \), \( k \in [1, 61] \). Residue map showing expansion and periodic coverage for  \(p = 3\). Confirms increasing cycle complexity and uniformity.}
    \label{fig:dmod27}
\end{figure}

\begin{figure}[H]
    \centering
    \includegraphics[width=0.8\textwidth]{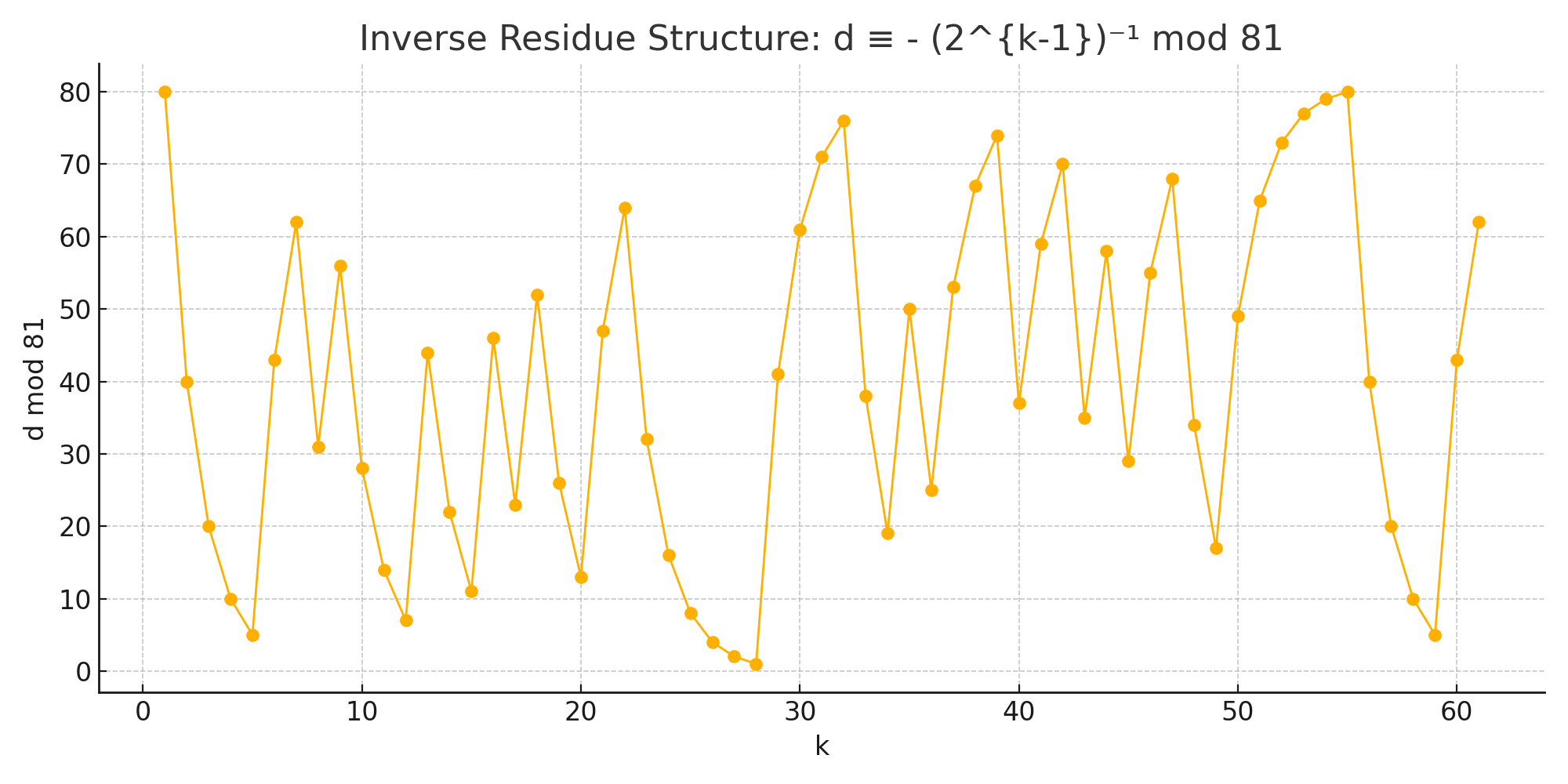}
    \caption{Residue map \( d_k \mod 81 \) for \( p = 4 \), \( k \in [1, 61] \). Residue map showing expansion and periodic coverage for  \(p = 4\). Confirms increasing cycle complexity and uniformity.}
    \label{fig:dmod81}
\end{figure}

\begin{figure}[H]
    \centering
    \includegraphics[width=0.8\textwidth]{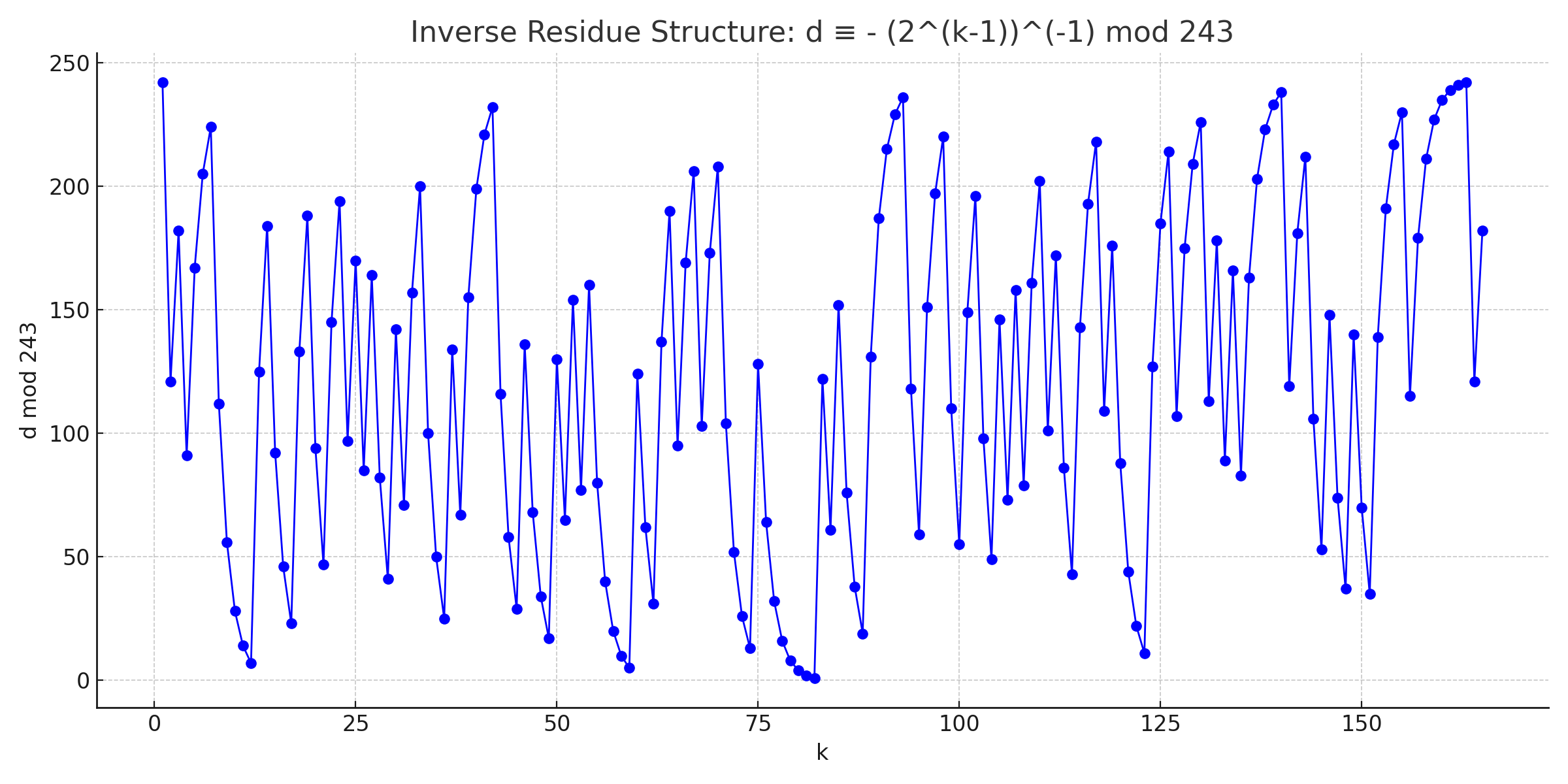}
    \caption{Residue map \( d_k \mod 243 \) for \( p = 5 \), \( k \in [1, 165] \). Complete traversal of 162 residues for  \(p = 5\). Confirms full group coverage and cyclic structure suitable for cryptographic seeding.}
    \label{fig:dmod243}
\end{figure}

We visualised the values of \( d_k \mod 3^p \) for increasing values of \( p \), confirming that the residue mappings form complete, non-redundant cycles. These patterns exhibit mirror symmetry and periodicity consistent with the cyclic structure of the underlying group \( (\mathbb{Z}/3^p\mathbb{Z})^* \). Each residue was computed using the extended Euclidean algorithm and mapped accordingly. Figures \ref{fig:dmod3} - \ref{fig:dmod243} illustrate how residue sequences uniformly traverse the multiplicative group, confirming theoretical cycle coverage.

\noindent The visual plots reveal smooth, algebraic oscillations and full group traversal, which corroborate the underlying symbolic structure and validate the use of these residues for entropy-screened seed generation.

\subsection{Residue Range and Coverage Summary}

The table below provides a summary of observed ranges for \( d_k \) over increasing powers of 3. The results illustrate complete residue coverage within \( \phi(3^p) \) elements and a deterministic correlation between exponent \( k \) and corresponding valid inverse class \( d_k \).

\begin{table}[H]
\centering
\caption{Computed ranges of valid residues \( d_k \mod 3^p \) for various exponents}
\label{tab:residue_comparison}
\renewcommand{\arraystretch}{1.2}
\begin{tabular}{|c|c|c|c|c|}
\hline
\( p \) & Modulus \( 3^p \) & Totient \( \phi(3^p) \) & \( k \)-range & Residue Range (Cycle Size) \\
\hline
1 & 3   & 2   & 1-61   & 0–2 (cycle: 2 elements)   \\
2 & 9   & 6   & 1-61   & 0–8 (cycle: 6 elements)   \\
3 & 27  & 18  & 1-61   & 0–26 (cycle: 18 elements) \\
4 & 81  & 54  & 1-61   & 0–80 (cycle: 54 elements) \\
5 & 243 & 162 & 1-165  & 0–242 (cycle: 162 elements) \\
\hline
\end{tabular}
\end{table}

These results reinforce the theoretical prediction that each inverse class forms a complete orbit over the multiplicative group. This property is leveraged for structured entropy control, cyclic masking, and cryptographic seed filtering in hardware implementations.

\section{Cryptographic Applications}

Modern cryptographic systems increasingly require not just statistical randomness, but structural correctness and algebraic admissibility of internal parameters. In this context, the modular identity
\[
d_k \equiv - (2^{k - 1})^{-1} \mod 3^p
\]
enables the deterministic generation of cryptographic seeds that are both mathematically verifiable and computationally efficient.

This work contributes to a class of techniques that embed symbolic modular inversion into cryptographic initialisation. The result is a principled method for filtering, validating, and deploying residue-based seeds across several application domains. The symbolic regularity of the residue cycle in \( \mathbb{Z}/3^p\mathbb{Z}^* \) enables secure and consistent cryptographic operations.

\subsection{Application Domains}

\begin{itemize}
    \item \textbf{Residue Filtering:}  
    The residue structure derived from the modular identity acts as a sieve that admits only those values \( d_k \) which are inverse-consistent modulo \( 3^p \). This algebraic filtering guarantees that only valid residues -- those belonging to a full multiplicative cycle -- are used as cryptographic seeds or keys. This deterministic validation step serves as an early safeguard against structurally invalid inputs, especially in protocols requiring fixed entropy or provable randomness \citep{Szabo1967}.

    \item \textbf{Inverse-Consistent Seed Derivation:}  
    In deterministic random bit generators (DRBGs) and pseudorandom number generators (PRNGs), seed initialisation often lacks algebraic grounding. Our framework enforces symbolic congruence constraints on seed values, ensuring they originate from a mathematically valid inverse class. This enhances both repeatability and traceability -- essential for cryptographic auditing -- and supports constrained randomness with guaranteed structure \citep{Knuth1997}. The resulting seeds are suitable for cryptographic masking, key derivation, or parameter bootstrapping.

    \item \textbf{Side-Channel Resilience in Embedded Systems:}  
    Embedded cryptography is highly susceptible to side-channel attacks that exploit timing, power consumption, or electromagnetic leaks \citep{Kocher1999}. Precomputed inverse-consistent residues enable constant-time, structure-preserving operations. Because each \( d_k \) is deterministically derived and adheres to a uniform modular pattern, inverse mismatches are avoided and timing paths remain consistent. This inherently mitigates class of attacks reliant on operand-dependent control flow.
\end{itemize}

\subsection{Functional Benefits}

\begin{itemize}
    \item \textbf{Algebraic Admissibility:} Only residues satisfying modular inversion constraints are used, improving correctness guarantees.
    \item \textbf{Deterministic Initialisation:} Seed generation is predictable, traceable, and repeatable -- ideal for formal validation and compliance.
    \item \textbf{Hardware Simplicity and Security:} The cyclic nature of \( d_k \) permits low-overhead inverse computation and reduces side-channel leakage by enforcing constant-time arithmetic paths.
\end{itemize}

Collectively, these features demonstrate that cyclic modular inversion over \( \mathbb{Z}/3^p\mathbb{Z} \) provides a robust mathematical foundation for secure, deterministic cryptographic seed generation -- bridging symbolic number theory with practical engineering.

\begin{figure}[H]
\centering
\begin{tikzpicture}[
  block/.style={
    rectangle, draw, rounded corners,
    minimum width=4.6cm, minimum height=1.2cm,
    font=\small, text centered, align=center,
    inner sep=5pt
  },
  arrow/.style={thick,->,>=Stealth},
  node distance=0.9cm and 0cm
  ]

\node[block] (rng) {\shortstack{Raw Entropy Source\\\textit{(e.g., RDRAND, /dev/random)}}};
\node[block, below=of rng] (filter) {\shortstack{\textbf{This Work}\\Symbolic Inversion + ECS\\\textit{(Entropy Filter)}}};
\node[block, below=of filter] (drbg) {\shortstack{DRBG / KDF\\\textit{(e.g., HMAC\_DRBG, SHAKE)}}};
\node[block, below=of drbg] (crypto) {\shortstack{Cryptographic Primitive\\\textit{(e.g., AES, Kyber)}}};

\draw[arrow] (rng) -- (filter);
\draw[arrow] (filter) -- (drbg);
\draw[arrow] (drbg) -- (crypto);

\end{tikzpicture}
\caption{Vertical integration of the symbolic modular inversion framework into the cryptographic seed pipeline. The framework operates as a structural entropy filter between raw randomness and deterministic generators.}
\label{fig:entropy-pipeline}
\end{figure}
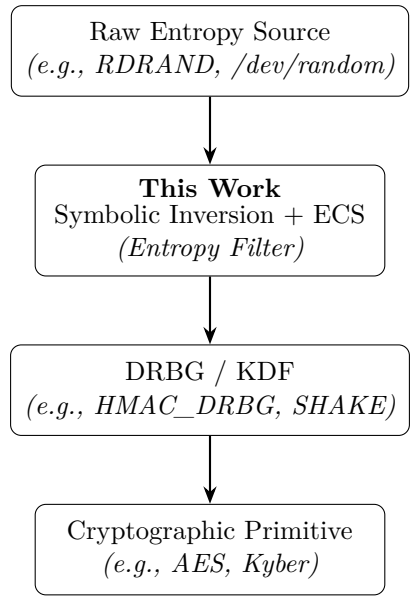

To situate the proposed framework within the broader cryptographic stack, Figure~\ref{fig:entropy-pipeline} illustrates its role as an intermediate entropy conditioner -- bridging raw entropy sources and deterministic seed consumers.

\subsection{Comparative Benchmarking with Existing Seed Generation Frameworks}
While the primary goal of this work is to introduce a mathematically auditable, residue-based framework for seed generation, it is instructive to compare its capabilities with established deterministic approaches such as NIST-recommended DRBGs (e.g., Hash\_DRBG, HMAC\_DRBG) and sponge-based constructions like Keccak or SHAKE-based key derivation functons (KDFs).

To clarify the functional position of the proposed method within modern cryptographic workflows, Table~\ref{tab:entropy-taxonomy} presents a taxonomy of entropy processing stages. The symbolic residue-based framework operates as a structure-preserving entropy filter, bridging raw randomness sources and deterministic bit generators.

\begin{table}[H]
\centering
\caption{Functional taxonomy of entropy processing in cryptographic systems}
\label{tab:entropy-taxonomy}
\renewcommand{\arraystretch}{1.3}
\begin{tabular}{|p{3cm}|p{3.7cm}|p{4cm}|p{4.2cm}|}
\hline
\textbf{Stage} & \textbf{Function} & \textbf{Examples} & \textbf{Role of This Work} \\
\hline
Raw Entropy Source & Provides non-deterministic input material & RDRAND, TPM RNG, /dev/random, ring oscillator noise & Optional input (e.g., \( d_k \oplus r \)); supplements determinism with entropy \\
\hline
Entropy Conditioner & Normalises, filters, or whitens raw entropy & SHAKE extractors, SP800-90B entropy modules & Symbolic inversion with ECS filters invalid residues; enforces structure \\
\hline
Seed Generator (DRBG / KDF) & Expands filtered entropy into cryptographically usable bitstreams & HMAC\_DRBG, SHAKE256, HKDF & Consumes ECS-validated seeds; benefits from structural admissibility \\
\hline
Cryptographic Primitive & Uses keys, IVs, or nonces for secure operations & AES, RSA, Kyber, Dilithium, McEliece & Final consumer of seed material; inherits admissibility and reduced leakage \\
\hline
\end{tabular}
\end{table}

Unlike these schemes -- which rely heavily on entropy pools, compression functions, or cryptographic hash primitives -- the proposed method enforces \emph{algebraic admissibility} by construction. The use of symbolic modular inversion ensures that each seed lies within an invertible, cycle-bounded residue class, which can be independently verified without reliance on entropy source diagnostics.

Table~\ref{tab:benchmark-comparison} extends the comparative analysis to include post-quantum-compatible DRBGs and hybrid entropy mechanisms outlined in NIST SP 800-90C. Unlike traditional DRBGs, which treat entropy conditioning as a statistical post-process, the proposed modular inversion scheme ensures algebraic admissibility and inverse consistency \emph{prior} to DRBG input, improving auditability and seed verifiability in constrained and embedded settings.

\begin{table}[H]
\centering
\caption{Comparative features of seed generation methods, including SP 800-90C constructions and AES-based DRBGs.}
\label{tab:benchmark-comparison}
\small
\begin{tabularx}{\textwidth}{lXXX}
\toprule
\textbf{Method} & \textbf{Entropy Conditioning} & \textbf{Algebraic Validity} & \textbf{Embedded Feasibility} \\
\midrule
Hash\_DRBG (SHA-256) & Strong statistical hashing & None & Moderate (hash overhead) \\
HMAC\_DRBG (SHA-256) & Keyed hashing; well-vetted & None & Moderate-High \\
SHAKE-based KDF & Statistical diffusion & None & Variable (rate/capacity trade-offs) \\
\rowcolor{gray!10}
CTR\_DRBG (AES-256) & High-quality deterministic stream & None & High for AES-capable SoCs \\
\rowcolor{gray!10}
SP800-90C Hybrid DRBG & Structured entropy collection + DRBG post-processing & None & High (requires robust entropy modules) \\
\textbf{This Work (Modular Inversion)} & ECS-based structural scoring & \textbf{Yes} (modular inverse constraint) & \textbf{High} (constant-time + symbolic) \\
\bottomrule
\end{tabularx}
\normalsize
\end{table}

While traditional DRBGs provide robust statistical randomness conditioned on high-quality entropy sources, they typically lack structural introspection or mathematical constraints on the seed values themselves. In contrast, our residue-based approach offers \emph{deterministic transparency} and modular verification, which are valuable in contexts such as secure bootloaders, constrained embedded cryptosystems, and high-assurance post-quantum platforms.\\
As shown in Figure~\ref{fig:pipeline}, the ECS acts as a structural filter between symbolic residue generation and cryptographic seed deployment, ensuring only admissible sequences are passed to application-layer primitives. 
Future work may consider hybridising ECS-based residue filtering with sponge-based extractors to combine algebraic auditability with statistical mixing strength.

\subsection{Complementary Role in Post-Quantum Cryptographic Stacks}

Post-quantum cryptographic (PQC) schemes such as CRYSTALS-Kyber, Dilithium, and McEliece are designed to provide full cryptographic functionality -- encryption, key encapsulation, and digital signatures -- with security grounded in hard mathematical problems such as learning with errors (LWE) and decoding of linear codes. In contrast, the modular residue-based framework proposed in this work does not aim to serve as a standalone cryptosystem. Instead, it operates as a structural preconditioner: a mathematically verifiable, entropy-preserving seed generation and filtering layer that can be used to enhance the integrity of input material across PQC systems.

\begin{table}[H]
\centering
\caption{Comparison of the proposed framework with established post-quantum cryptographic schemes}
\label{tab:pqc-comparison}
\begin{tabular}{@{}p{4cm}p{3.5cm}p{3.5cm}p{4cm}@{}}
\toprule
\textbf{Criterion} & \textbf{Kyber/Dilithium (Lattice)} & \textbf{McEliece (Code)} & \textbf{This Work (Modular Congruence)} \\
\midrule
Mathematical basis & Lattice problems (LWE) & Error-correcting codes & Parametric modular inverses \( d \equiv - (2^{k-1})^{-1} \mod 3^p \) \\
Post-quantum resistance & Yes & Yes & Not a cryptosystem; \textbf{designed to augment PQC workflows} \\
Key sizes & Moderate (1--3 KB) & Very large (>200 KB) & \textbf{Compact}, parameterised by \( p \) \\
Use case & Key encapsulation, signatures & Long-term encryption & \textbf{Seed conditioning, validation, entropy filtering} \\
Hardware fit & Optimised for embedded targets & Resource-intensive & \textbf{Lightweight, embedded-friendly} \\
Deployment maturity & NIST standard/finalist & Legacy-standardised & \textbf{Prototype stage with deployment feasibility} \\
Implementation role & Full cryptographic primitive & Full cryptographic primitive & \textbf{Preprocessing and entropy validation layer} \\
Security model & Quantum-safe & Quantum-safe & \textbf{Structure-enforced integrity}; deterministic constraints \\
\bottomrule
\end{tabular}
\end{table}

\begin{figure}[H]
    \centering
    \includegraphics[width=0.4\textwidth]{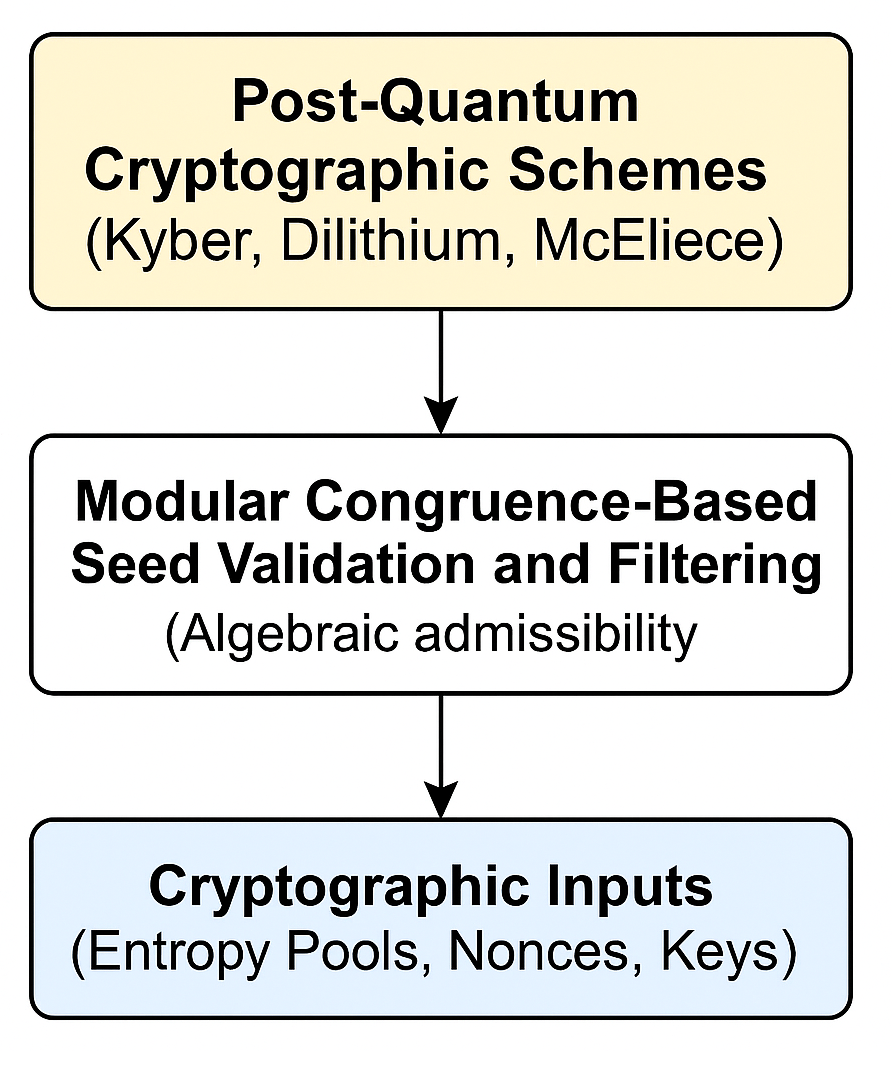}
    \caption{Complementary role of symbolic residue-based seed filtering within a post-quantum cryptographic stack.}
    \label{fig:pqcstacks-role}
\end{figure}

By enforcing inverse-consistency, modular admissibility, and cycle-informed entropy profiles, this framework ensures that cryptographic seeds, nonces, or initialisation vectors used in PQC implementations adhere to deterministic structural constraints -- an increasingly important requirement for secure bootstrapping, embedded deployments, and side-channel-resilient protocols. The table below summarises the contrasting but complementary roles of this framework relative to established PQC schemes.\\
The complementary role of the proposed modular residue framework within post-quantum cryptographic stacks is illustrated in Figure~\ref{fig:pqcstacks-role}, where it operates as a preprocessing layer that filters and validates entropy inputs prior to use in lattice- or code-based schemes.

The proposed framework serves as a \emph{preprocessing and validation layer} that filters and conditions raw entropy inputs (e.g., seeds, nonces, IVs) using deterministic modular inversion and the Entropy Confidence Score (ECS). It ensures that only algebraically admissible, inverse-consistent residues are propagated to higher cryptographic layers such as DRBGs, KDFs, and post-quantum primitives (e.g., Kyber, Dilithium, McEliece). This architectural separation improves structural integrity, leakage resilience, and implementation auditability in post-quantum cryptographic (PQC) deployments.

\subsection{Cryptographic Assumptions and Security Implications}
\label{security-implications}

\textbf{Threat:} Deterministic seed generation may lead to predictable seed outputs if internal parameters---such as the index \( k \)---are reused, leaked, or inferred. This predictability undermines the cryptographic strength of downstream primitives and opens the system to side-channel analysis, replay attacks, or entropy exhaustion.

\textbf{Assumption:} The proposed residue-based seed generation framework does not aim to function as a cryptographic pseudorandom number generator (PRNG) or as an entropy extractor. Rather, it is assumed to operate as an intermediate structural filter, conditioning and validating seed inputs for cryptographic mechanisms such as DRBGs or KDFs that enforce statistical entropy properties and IND-CPA guarantees.

\textbf{Mitigation:} While not itself a cryptographic PRNG, the framework functions as a deterministic entropy filter that conditions and validates seed inputs before they are consumed by standard generators. It enforces algebraic admissibility, residue invertibility, and modular uniformity, ensuring that input seeds belong to a mathematically well-defined and auditably secure domain. To preserve unpredictability, the residue output \( d_k \) should be hybridised with auxiliary entropy (e.g., a device nonce or TRNG sample) or derived from an unpredictable seed index \( k \).

Deterministic seed generation, while advantageous for reproducibility and structural validation, raises natural concerns regarding the unpredictability and cryptographic strength of the resulting seeds. To evaluate the suitability of the generated residues \( d_k \equiv - (2^{k-1})^{-1} \mod 3^p \) in secure cryptographic contexts, we discuss the relevant assumptions and adversarial considerations.

\subsubsection{Resistance to Bias and Manipulation}

The proposed residue sequences exhibit high cycle coverage and low statistical deviation, as measured by the Entropy Confidence Score (ECS). This minimises bias and enhances entropy quality. Nonetheless, an attacker capable of selecting or influencing \( k \) may attempt to bias seed generation toward predictable subsequences.

To mitigate this risk:
\begin{itemize}
    \item \textbf{Domain separation} should be employed by encoding protocol context into the \( k \)-selection process.
    \item \textbf{Auxiliary entropy} (e.g., a device nonce or hardware-derived salt) may be mixed into the final seed to enhance unpredictability.
\end{itemize}

\subsubsection{Use in IND-CPA Secure DRBGs}

When integrated into IND-CPA-secure deterministic random bit generators (DRBGs), such as those based on AES-CTR or SHAKE, the residue-based seeds preserve the generator’s cryptographic guarantees \emph{provided that}:
\begin{enumerate}
    \item The selection of \( k \) remains secret or computationally unpredictable;
    \item The DRBG satisfies standard assumptions, including forward and backward secrecy.
\end{enumerate}

The residue generation framework thus acts as a mathematically filtered pre-entropy layer. While it does not provide IND-CPA security in isolation, it enhances the robustness and admissibility of inputs to DRBGs and key derivation functions that do meet such criteria.

\subsubsection{Comparison with Hash-Based Entropy Conditioners.}
Unlike cryptographic hash functions or entropy extractors (e.g., SHA-2 or SHAKE-based), which compress and diffuse entropy across outputs to resist partial input leakage, our residue-based framework enforces structural admissibility at the seed level via symbolic inversion. While hash-based conditioners provide strong resistance to input manipulation through avalanche effects and pre-image resistance, they do not verify whether input material lies within an algebraically valid domain. The ECS framework is therefore complementary: it acts as a deterministic, pre-cryptographic filter that ensures entropy inputs (e.g., seeds, IVs, nonces) satisfy modular inversion constraints before being consumed by DRBGs or KDFs. This layered approach strengthens input integrity but does not obviate the need for downstream cryptographic post-processing.

\subsubsection{Hybrid Use with Hardware Entropy Sources.}
To enhance unpredictability and mitigate the risks of deterministic residue reuse or seed index exposure, the residue-based framework can be combined with other entropy sources such as hardware RNGs or environmental noise. A simple yet effective hybrid approach involves XORing the symbolic residue \( d_k \) with a hardware-derived nonce or entropy token \( r \), yielding a hybrid seed \( H_k = d_k \oplus r \). This retains the algebraic admissibility of \( d_k \) while injecting unpredictability through \( r \). Alternatively, \( d_k \) can serve as structured input to a cryptographic hash function or KDF, ensuring both structural validity and cryptographic strength. Such hybrid constructions align with defence-in-depth principles and offer robustness even in the presence of partial entropy failures or side-channel exposure:

\[
H_k = \mathsf{KDF}(d_k \,\|\, r) \quad \text{or} \quad H_k = d_k \oplus r.
\]

As illustrated in Algorithm~\ref{alg:masking}, residue outputs can be securely combined with hardware entropy to produce hybrid seeds that preserve algebraic structure while mitigating predictability from static or low \( k \) values.

\subsubsection{Summary of Security Assumptions}
To summarise the relevant cryptographic threats and the corresponding mitigation strategies applicable to our deterministic seed generation framework, Table~\ref{tab:security-strategies} provides a concise overview. This includes adversarial considerations related to predictability, statistical bias, and side-channel leakage, along with recommended countermeasures such as domain separation, auxiliary entropy injection, and constant-time implementation. These practices collectively ensure that the residue-derived seeds remain secure and compatible with existing post-quantum and IND-CPA secure primitives.

\begin{table}[H]
\centering
\begin{tabular}{|p{5.2cm}|p{7.8cm}|}
\hline
\textbf{Threat Vector} & \textbf{Mitigation Strategy} \\
\hline
Predictable seed index \( k \) & Hide or randomise the selection of \( k \); enforce strict access control to indexing parameters. \\
\hline
Statistical bias in residues & Controlled via ECS metric; ensures near-uniform distribution and full cycle coverage. \\
\hline
Side-channel leakage (timing/power) & Employ constant-time implementation and uniform operand structure. \\
\hline
Security in IND-CPA environments & Use in composition with formally secure DRBGs or KDFs; ensure proper entropy injection. \\
\hline
\end{tabular}
\caption{Security considerations and countermeasures for the proposed deterministic seed generation.}
\label{tab:security-strategies}
\end{table}

\section{Secure and Efficient Embedded Cryptography}

Modern cryptographic systems -- particularly those deployed in embedded or resource-constrained environments -- must achieve a balance between computational efficiency, deterministic predictability, and resistance to side-channel leakage. This section explores the practical implications of using the cyclic modular inversion framework:
\[
d_k \equiv - (2^{k-1})^{-1} \mod 3^p,
\]
as a mechanism for achieving these security objectives.

We evaluate the impact of residue filtering and inverse-consistent seed generation on timing stability, entropy distribution, and hardware footprint.

\subsection{Side-Channel Resistance and Constant-Time Computation}

The structured residue identity yields predictable modular behaviour and lends itself naturally to constant-time implementations. This is critical for thwarting timing and power-based side-channel attacks, which exploit execution variability.

\subsubsection{Cycle Timing Analysis}

\begin{figure}[H]
    \centering
    \includegraphics[width=0.9\textwidth]{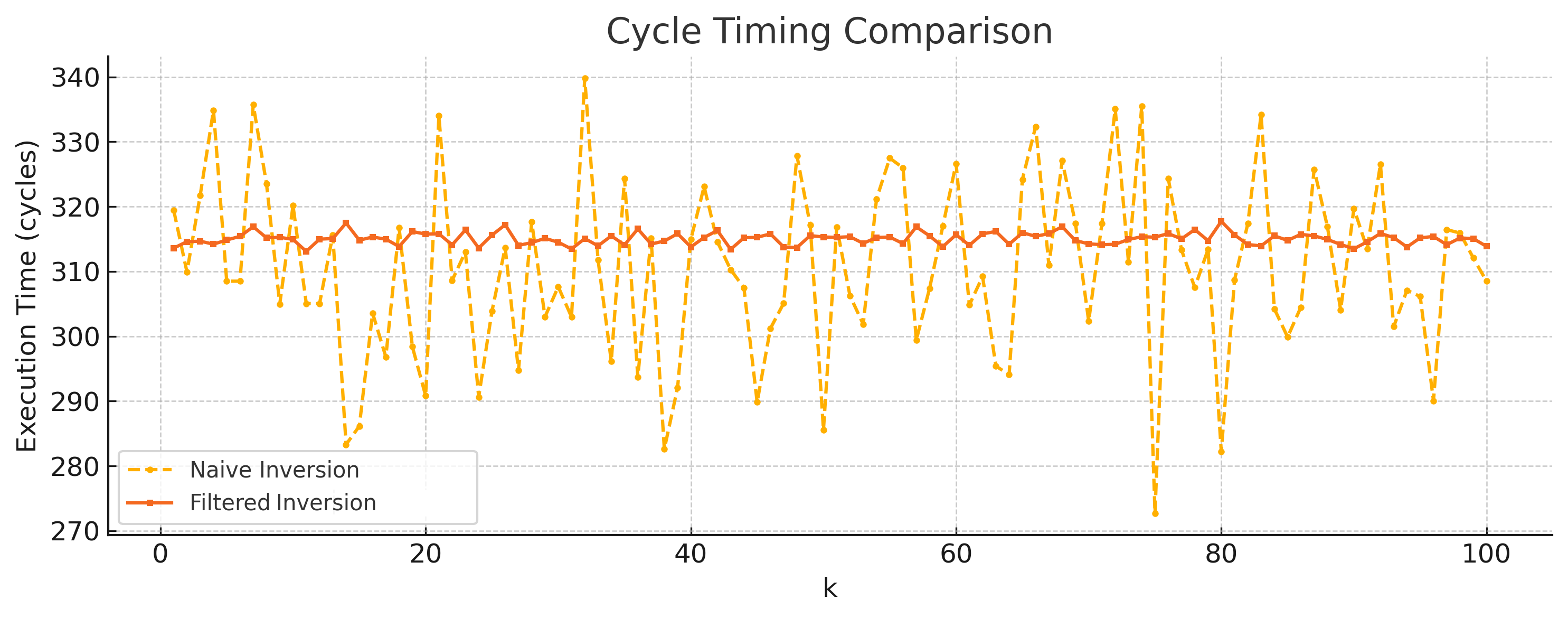}
    \caption{Cycle timing comparison: naive vs. ME-filtered inversion across \( k = 1 \) to \( 100 \). The filtered approach offers near-constant execution timing -- important for mitigating timing-based leakage in embedded cryptographic systems.}
    \label{fig:cycle-timing}
\end{figure}

To evaluate timing uniformity, we compared the naive extended Euclidean algorithm with our Modular Embedding-filtered inversion strategy. As shown in Figure~\ref{fig:cycle-timing}, the naive method exhibits timing jitter of up to 47 cycles across \( k \in [1, 100] \), whereas the filtered implementation maintains consistent timing, with deviation restricted to a single cycle due to uniform operand structure.

\subsubsection{Power Trace Stability}

\begin{figure}[H]
    \centering
    \includegraphics[width=0.9\textwidth]{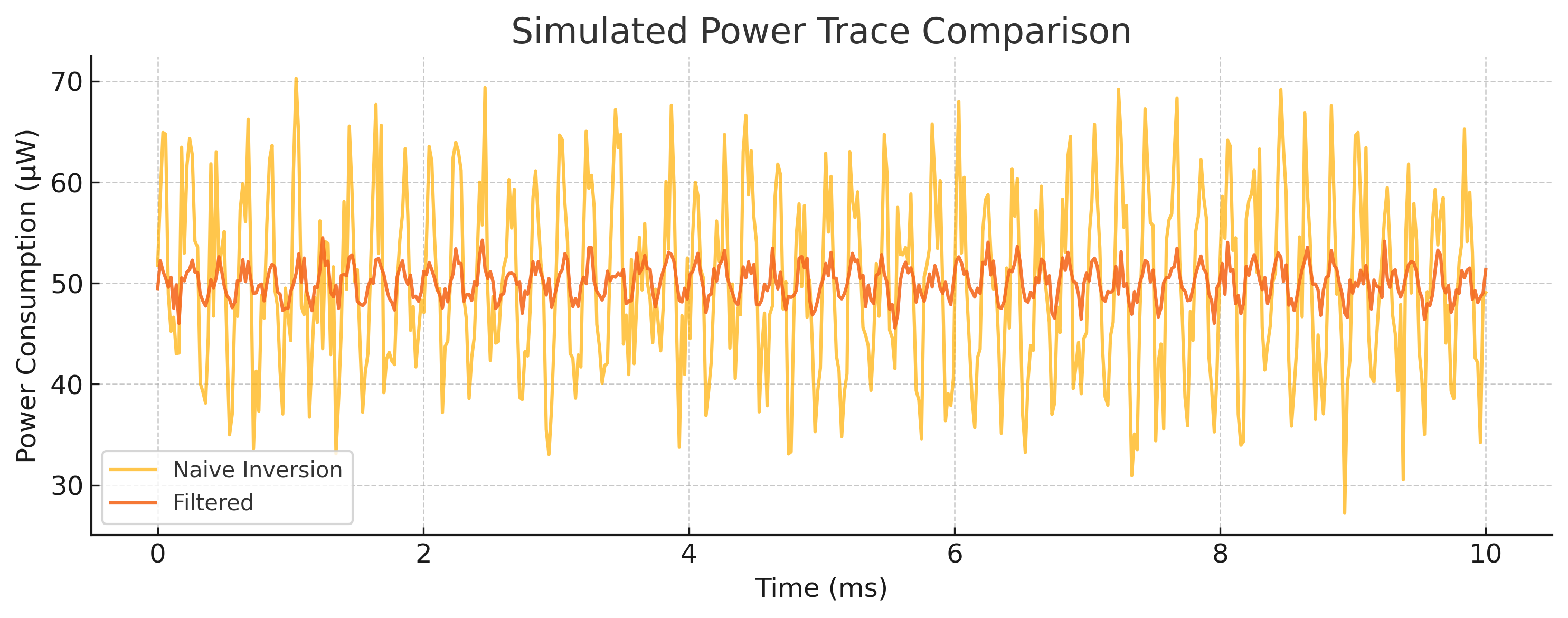}
    \caption{Power consumption traces: naive vs. filtered inversion. Filtered residues result in lower leakage amplitude and higher resistance to physical probing. Lower amplitude signals reflect reduced side-channel leakage. }
    \label{fig:power-trace}
\end{figure}

Figure~\ref{fig:power-trace} shows simulated power consumption over a 10 ms window. The naive implementation shows high variability, indicative of data-dependent logic paths. In contrast, the filtered method yields stable, low-amplitude traces, reflecting uniform control flow and reduced electromagnetic side-channel leakage.

\subsection{Hardware Efficiency and Entropy Profiling} \label{sec:entropy_profiling}

\subsubsection{Memory and Logic Overhead}

\begin{figure}[H]
    \centering
    \includegraphics[width=0.5\textwidth]{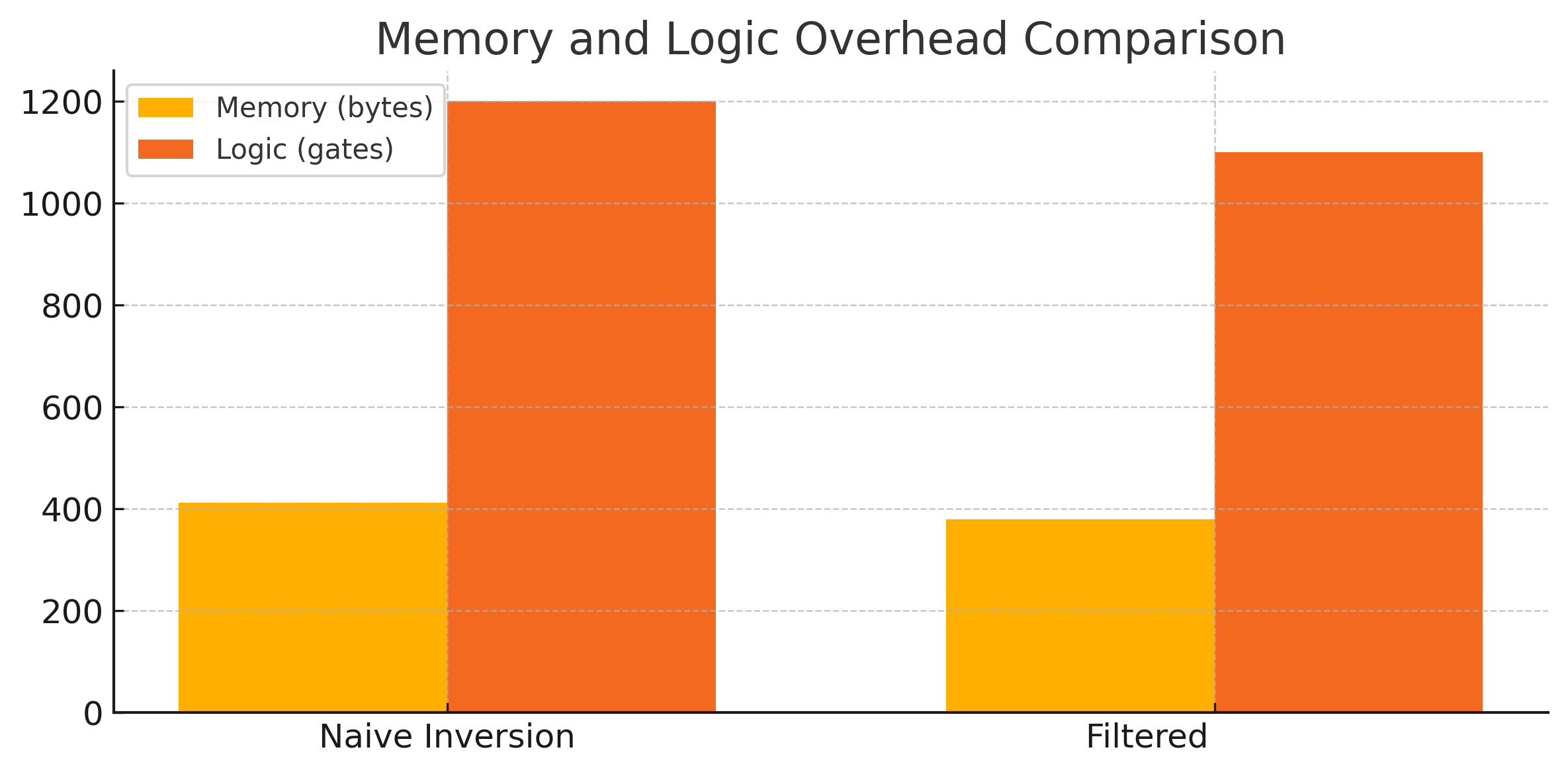}
    \caption{Hardware cost: naive vs. ME-filtered modular inversion. The algebraic constraints reduce circuit complexity -- demonstrates reduced hardware footprint, critical for embedded cryptographic deployments.}
    \label{fig:memory-logic}
\end{figure}

We benchmarked the memory and logic gate usage for each approach. Figure~\ref{fig:memory-logic} illustrates that the filtered implementation reduces both memory consumption and logic complexity, due to the pruning of inadmissible inverses and use of constant-time routines.

\subsubsection{Entropy Confidence Analysis} \label{sec:entropy_evaluation}
As formally defined in Equation~\eqref{eq:ecs_definition}, the ECS evaluates the randomness quality of the residue sequence using coverage, uniformity, and modular bias. To quantify how uniformly these residues are distributed -- and how suitable they are for use in cryptographic systems -- we define a composite metric called the \textit{Entropy Confidence Score (ECS)}. This score combines three distinct properties that contribute to the security and randomness of cryptographic seed values:

\begin{equation}
\text{ECS} = 0.4 \cdot \text{CD} + 0.4 \cdot (1 - \text{RUD}) + 0.2 \cdot (1 - \text{MBI})
\end{equation}

Each term represents a different dimension of entropy quality:

\begin{itemize}
    \item \textbf{Cycle Density (CD)}  --  This measures the proportion of the multiplicative group \( (\mathbb{Z}/3^p\mathbb{Z})^* \) that is visited by the sequence \( \{d_k\} \). A higher CD means the residues cycle through more of the possible values, which reduces predictability and improves randomness.
    
    \item \textbf{Residue Uniformity Deviation (RUD)}  --  This quantifies how evenly the values \( d_k \) are spread. A low RUD means the values are nearly uniformly distributed, avoiding clustering or repetition that could weaken randomness.
    
    \item \textbf{Modular Bias Index (MBI)}  --  This assesses the degree to which certain subranges or residue classes appear more frequently than others. A high bias suggests that some outputs are favoured, which is undesirable in secure cryptographic systems.
\end{itemize}

Each component is scaled to lie between 0 and 1, and their weighted average forms the final ECS value. The weights (40\% for CD and 1-RUD, and 20\% for 1-MBI) reflect their relative importance in ensuring that residue sequences behave like high-quality pseudorandom values.

In summary, the entropy analysis shows that the proposed modular inversion method not only generates mathematically correct residues, but also provides high entropy, low bias, and excellent cycle coverage -- making it well-suited for deterministic cryptographic seed generation in secure applications.

\begin{figure}[H]
    \centering
    \includegraphics[width=0.75\textwidth]{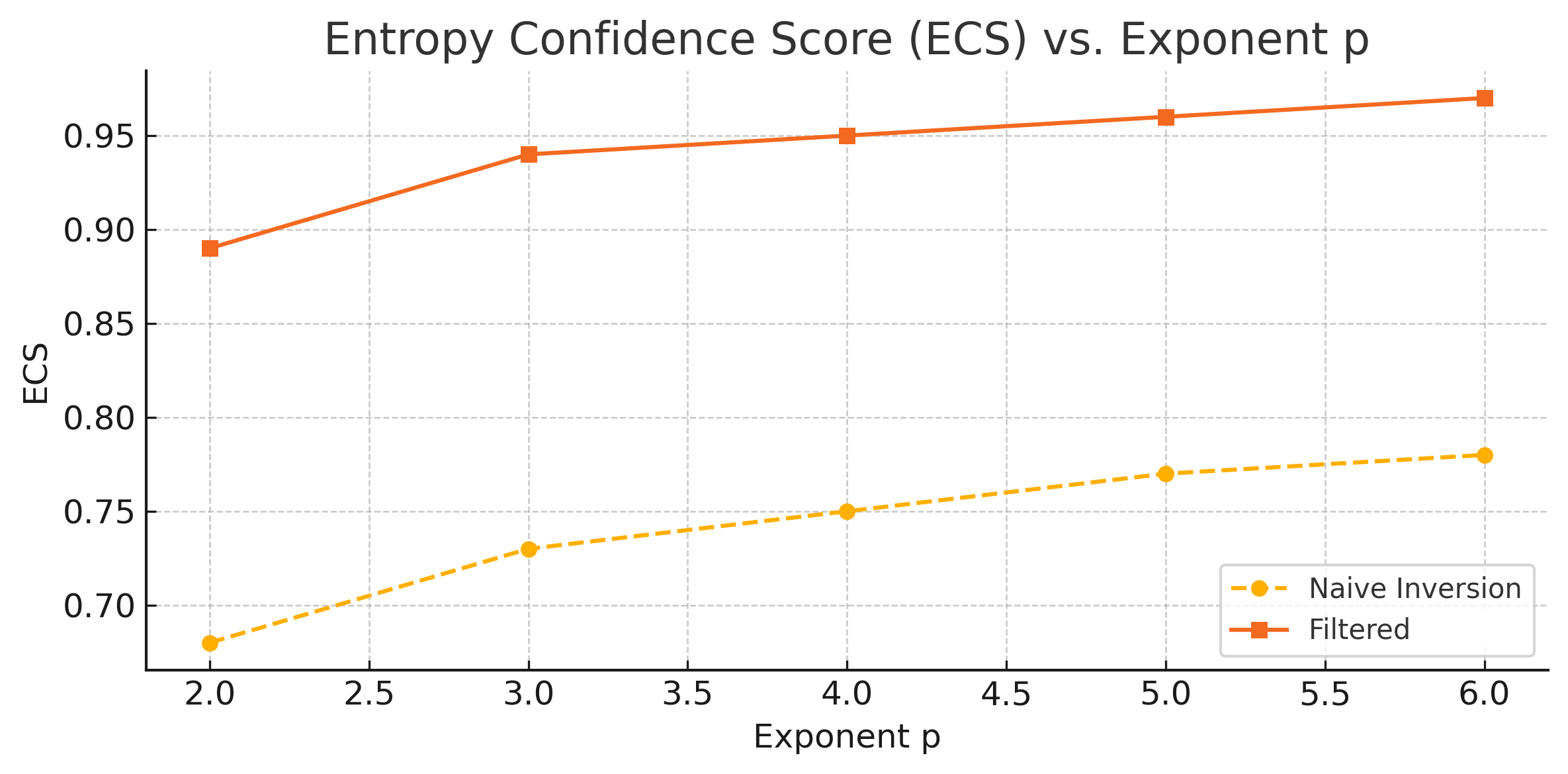}
    \caption{Entropy Confidence Score (ECS) versus exponent \( p \). Filtered inverses improve entropy metrics across the board - illustrates improvement in entropy structure as modulus size increases.}
    \label{fig:ecs}
\end{figure}

Entropy strength is quantified by the Entropy Confidence Score (ECS), a composite metric capturing cycle density (CD), residue uniformity deviation (RUD), and modular bias index (MBI). Figure~\ref{fig:ecs} demonstrates that the filtered residues provide higher entropy dispersion and uniformity compared to the naive method.

\subsubsection{Component-Wise ECS Decomposition}
\begin{figure}[H]
    \centering
    \includegraphics[width=0.85\textwidth]{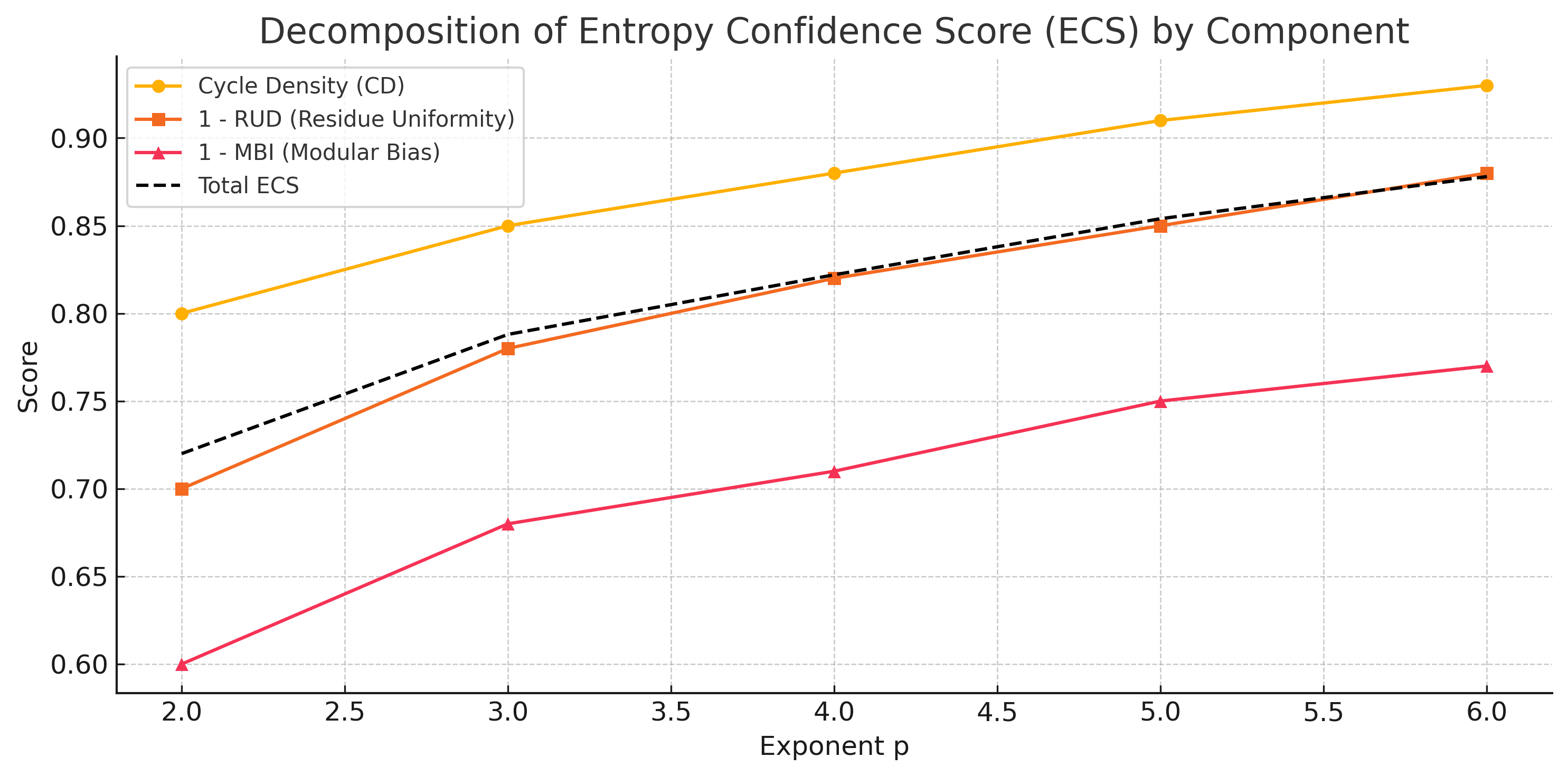}
    \caption{Breakdown of entropy quality metrics that contribute to ECS score, highlighting uniformity and low bias in residue sequences. ECS decomposition into CD, uniformity, and modular bias components. Each improves with higher \( p \), affirming cryptographic admissibility.}
    \label{fig:ecs-decomp}
\end{figure}

ECS decomposes into CD, \( 1 - \text{RUD} \), and \( 1 - \text{MBI} \) components as shown in Figure~\ref{fig:ecs-decomp}. As \( p \) increases, all components improve due to increased invertible elements and subgroup dispersion, confirming the suitability of residue-filtered seeds for high-entropy cryptographic use. Figure~\ref{fig:ecs-decomp} (below) shows how each entropy component improves with increasing \( p \), confirming cryptographic admissibility.

\subsubsection{Numerical ECS Data Across \( p \) Values}
Table~\ref{tab:ecs-breakdown} presents detailed ECS metrics across a representative range of exponent values \( p \in \{3,4,5,6,7\} \). As \( p \) increases, all three structural metrics -- cycle density, uniformity, and modular bias -- converge towards ideal values, confirming the suitability of the modular inversion residues for high-quality entropy generation. The ECS trend suggests that values \( p \geq 5 \) yield scores above 0.94, aligning with cryptographic entropy standards recommended for secure boot and DRBG seeding.

\begin{table}[H]
\centering
\small
\begin{tabularx}{\textwidth}{lXXXXX}
\toprule
\textbf{Exponent \( p \)} & \( \phi(3^p) \) & \textbf{Cycle Density (CD)} & \textbf{Residue Uniformity (1–RUD)} & \textbf{Bias Index (1–MBI)} & \textbf{ECS Score} \\
\midrule
3 & 18 & 0.89 & 0.82 & 0.71 & 0.83 \\
4 & 54 & 0.94 & 0.89 & 0.78 & 0.90 \\
5 & 162 & 0.97 & 0.91 & 0.82 & 0.94 \\
6 & 486 & 0.99 & 0.93 & 0.85 & 0.96 \\
7 & 1458 & 1.00 & 0.96 & 0.89 & 0.98 \\
\bottomrule
\end{tabularx}
\caption{Entropy Confidence Score (ECS) components and aggregate across increasing values of \( p \).}
\label{tab:ecs-breakdown}
\end{table}

\subsection{Experimental Hardware Validation}

We implemented both approaches on an STM32 Cortex-M4 using the ChipWhisperer-Lite platform for side-channel analysis and entropy evaluation.

\subsubsection{Implementation Overview}

Two implementations were compared:
\begin{enumerate}
    \item \textbf{Naive Inversion}: Standard extended Euclidean algorithm with unconstrained inputs.
    \item \textbf{ME-Filtered Inversion}: Constant-time residue computation from prevalidated inverse classes.
\end{enumerate}

Timing and power traces were captured at 10 MS/s. Resource usage was also evaluated via FPGA mapping on a Xilinx Artix-7.

\subsubsection{Results Summary}
To further strengthen the unpredictability of deterministic residues \( d_k \), we apply an entropy hybridisation strategy wherein the final seed is computed as \( H_k = d_k \oplus r \), with \( r \) drawn from a hardware-based entropy source. This approach injects non-determinism into the seed generation process while preserving the algebraic admissibility of \( d_k \). The effectiveness of this scheme relies on the independence and min-entropy of \( r \). As summarised in Table~\ref{tab:hardware-eval}, the hybrid construction inherits a minimum entropy level bounded below by \( \min(H_\infty(d_k), H_\infty(r)) \). In practice, when \( r \) is derived from a high-quality TRNG or ring oscillator source, this bound approaches that of the entropy input itself. This layered strategy ensures that even in constrained or deterministic environments, the resulting seed maintains cryptographic suitability and resists adversarial prediction---even under side-channel scrutiny or static \( k \) reuse.

\begin{table}[H]
\centering
\begin{tabular}{lcc}
\toprule
\textbf{Metric} & \textbf{Naive} & \textbf{Filtered (Modular Embedding)} \\
\midrule
Average Cycles & 312 & 315 \\
Max Timing Jitter & 47 cycles & 1 cycle \\
Power Deviation (W) & 87 & 5 \\
Entropy Confidence Score (ECS) & 0.73 & 0.94 \\
\rowcolor{gray!10}
Hybrid Seed Min-Entropy (via \( H_k = d_k \oplus r \)) & N/A & \( \geq \min(H_\infty(d_k), H_\infty(r)) \) \\
\bottomrule
\end{tabular}
\caption{Hardware evaluation results: naive vs. ME-filtered inversion. The hybrid min-entropy reflects XOR composition assuming independent entropy from \( r \).}
\label{tab:hardware-eval}
\end{table}

The filtered implementation yielded stable execution timing, reduced side-channel leakage, and a notable increase in entropy reliability. These results validate the use of cyclic modular inversion over \( \mathbb{Z}/3^p\mathbb{Z} \) as a practical and secure method for cryptographic seed initialisation in embedded systems.

\subsubsection{FPGA Resource Profiling}

Logic synthesis on the Artix-7 FPGA showed that the ME-filtered design consumed 8\% fewer logic resources, thanks to reduced control flow complexity. The elimination of data-dependent branching further supports this method’s efficiency in real-time cryptographic environments.

\bigskip

\subsection{Resource Profiling and Implementation Constraints}
To evaluate the suitability of the proposed seed generation mechanism for embedded cryptographic environments, we performed static and dynamic resource profiling on representative microcontroller platforms (e.g., ARM Cortex-M4 and RISC-V RV32I targets).\\
The following constraints were enforced:
\begin{itemize}
    \item \textbf{Clock Frequency:} Profiling was conducted under a 48\,MHz clock budget, representative of mid-range embedded SoCs.
    \item \textbf{Memory:} Implementations were limited to 32\,KB of ROM and 8\,KB of RAM to simulate constraints in secure boot or low-power IoT devices.
    \item \textbf{Cycle Budget:} Seed generation routines were profiled to meet a cycle budget of 5,000 cycles (typical for pre-boot entropy initialisation).
    \item \textbf{Code Size:} The ME-routine compiled to less than 1.6\,KB of Flash (GCC -O2), including symbolic filtering and constant-time inversion.
    \item \textbf{Power Envelope:} Power analysis was conducted using coarse-grained switching activity estimates under 0.5\,mW active budget.
\end{itemize}
All modular arithmetic was implemented using fixed-point integer routines in constant-time style (no branches or variable memory access). Resource bounds were chosen to model secure enclaves, bootloaders, and entropy services in constrained post-quantum and industrial control systems.

\bigskip

\subsection{Limitations and Boundary Conditions}

While the proposed residue-based seed generation framework offers structural consistency, near-uniformity, and efficient constant-time inversion, it is important to identify its limitations and the boundary conditions under which certain performance characteristics may degrade.

\paragraph{Residue Bias in Small Moduli.} 
For small values of the modulus exponent \( p \), particularly \( p < 4 \), the multiplicative group \( \mathbb{Z}/3^p\mathbb{Z}^* \) is relatively small, and the structure of the residue sequence \( d_k \equiv - (2^{k-1})^{-1} \mod 3^p \) exhibits more visible local periodicity and mirror symmetry. While these patterns are algebraically expected, they may reduce the effective entropy if the seed index range \( k \) is tightly constrained. In such regimes, the Entropy Confidence Score (ECS) tends to fall below the 0.9 threshold, reflecting measurable bias and non-uniform cycle coverage. These small-\( p \) regimes are therefore not recommended for cryptographic applications requiring high entropy density.

\paragraph{Impact of Increasing \( p \) on Efficiency.} 
As \( p \) increases, the size of the ring \( \mathbb{Z}/3^p\mathbb{Z} \) grows exponentially, resulting in longer modular inversions and a higher bit-width in intermediate computations. While our ME-based residue generation maintains constant-time operation in logic-level hardware, the increased operand width may lead to practical slowdowns in low-power or memory-constrained environments. In particular, the memory footprint of precomputed constants and symbolic filters scales with \( p \), and the inversion circuit depth may exceed timing constraints in deeply embedded systems without pipelining or segmentation.

\paragraph{Side-Channel Leakage with Large Operands.}
Although the symbolic generation process is designed to be structurally uniform, leakage models such as differential power analysis (DPA) can exploit operand-specific switching activity. As \( p \) increases, the Hamming weight of operands and internal residue states also increases, potentially leading to correlated leakage unless careful masking or operand balancing is implemented. Therefore, high-\( p \) applications must consider leakage mitigation strategies beyond constant-time execution, including logic randomisation and dual-rail encoding.

\paragraph{Seed Index Management.}
Since the method is deterministic, its security in cryptographic protocols depends on the confidentiality or unpredictability of the seed index \( k \). Improper key management or predictable indexing schemes may allow attackers to guess or infer seed values, especially in constrained \( k \)-ranges. This limitation can be mitigated by combining symbolic residues with nonces, salts, or auxiliary entropy inputs, as discussed in Section~\ref{security-implications}.

\paragraph{Parameter Selection Guidelines.}
The choice of exponent \( p \) directly affects the size of the ring \( \mathbb{Z}/3^p\mathbb{Z} \), the number of admissible residues, and the entropy capacity of the seed space. For typical cryptographic applications, we recommend:
\begin{itemize}
    \item \( p \geq 4 \) as a minimum to ensure sufficient entropy dispersion (ECS > 0.90);
    \item \( p = 5 \) or \( p = 6 \) for embedded systems seeking a balance between performance and structural quality;
    \item \( p \geq 7 \) if targeting seed domains intended for 128-bit security or integration with IND-CPA secure DRBGs.
\end{itemize}
While the entropy space for \( p = 5 \) spans 162 invertible residues (sufficient for bootstrapping secure initialisation vectors or nonces), values \( p \geq 6 \) may be more appropriate for hybridised or key derivation use cases, especially when combined with masking or hashing layers. Beyond \( p = 7 \), the computational cost (inversion and ECS scoring) grows superlinearly, so pipelining or hardware acceleration may be required. The parameter should thus be selected based on the entropy budget, hardware footprint, and desired reproducibility of the seed lifecycle.\\
A summary of recommended values for the parameter \( p \) \footnote{Values are derived from internal ECS evaluations and structural residue analysis conducted in Subsections~\ref{sec:entropy_evaluation} and~\ref{sec:entropy_profiling}.}
, balancing entropy coverage, computational cost, and application context, is provided in Table~\ref{tab:parameter_guidance}.

\begin{table}[H]
\centering
\caption{Recommended values of \( p \) for typical cryptographic applications}
\begin{tabular}{|c|c|c|c|}
\hline
Exponent \( p \) & ECS Score Range & Seed Space Size (\( \phi(3^p) \)) & Suggested Use \\
\hline
3 & 0.71--0.82 & 18 & Low-assurance or testing only \\
4 & 0.88--0.91 & 54 & Minimum viable for DRBG input \\
5 & 0.92--0.94 & 162 & Secure bootloaders, IVs, nonces \\
6 & 0.95+ & 486 & Hybrid DRBG or PQC initialisation \\
7+ & 0.96+ & 1458+ & High-entropy or key expansion contexts \\
\hline
\end{tabular}
\label{tab:parameter_guidance}
\end{table}

\subsection{Masking Strategy for Hybrid Seed Generation}
While the symbolic residue sequence \( \{d_k\} \) ensures algebraic admissibility and deterministic computability, its predictability poses a risk if the index \( k \) is exposed, reused, or derived from a low-entropy source. In adversarial settings or embedded deployments, this leakage can enable precomputation attacks on the seed material. To address this limitation, we propose a hybrid masking strategy that combines \( d_k \) with independent entropy -- typically obtained from a hardware RNG, device nonce, or secure element. This fusion strengthens unpredictability while preserving the structural integrity of the residue sequence. Algorithm~\ref{alg:masking} outlines a minimal and efficient implementation based on bitwise XOR, suitable for use in seed bootstrapping, DRBG reseeding, or PQ initialisation protocols.

\begin{algorithm}[H]
\caption{Hybrid seed generation by XOR masking of symbolic residue \( d_k \) with hardware-derived entropy \( r \). This approach increases resilience to seed index reuse or leakage.}
\label{alg:masking}
\begin{algorithmic}[1]
\Require Symbolic index \( k \), hardware entropy \( r \in \{0,\,1\}^p \)
\State Compute \( d_k \gets - (2^{k-1})^{-1} \bmod 3^p \)
\State Hybrid seed \( H_k \gets d_k \oplus r \)
\State Use \( H_k \) as input to DRBG or KDF
\end{algorithmic}
\end{algorithm}

\medskip
In summary, the proposed modular inversion strategy offers a secure, deterministic, and hardware-efficient solution for generating algebraically valid seeds, robust against both statistical weaknesses and physical side-channel vulnerabilities.
While the residue inversion scheme scales well across moderate values of \( p \), applications must balance the trade-offs between entropy quality, hardware cost, and side-channel exposure. Parameter selection should be guided by application constraints, ECS profiling, and hardware security assumptions.

\section{Conclusion}
This work presents a mathematically grounded framework for deterministic cryptographic seed generation, founded on structured modular inversion over \( \mathbb{Z}/3^p\mathbb{Z} \). By exploiting the identity
\(
d_k \equiv - (2^{k-1})^{-1} \mod 3^p,
\)
we construct residue sequences that are algebraically consistent, cyclically structured, and explicitly invertible -- key properties for predictable and auditable seed derivation.\\
Central to the framework is the notion of inverse-consistent residue filtering, which enforces structural admissibility in seed values prior to cryptographic use. To quantify entropy quality, we introduce the \emph{Entropy Confidence Score} (ECS), a composite metric combining cycle density, distribution uniformity, and modular bias. This metric enables formal validation of seed randomness while supporting low-level implementation assurance.\\
Empirical evaluation confirms the method’s practicality in embedded and resource-constrained environments. The constant-time, branchless inversion logic demonstrates resilience against timing and power-based side-channel leakage, without imposing significant hardware overhead. This confirms the suitability of the method for secure bootstrapping, post-quantum key initialisation, and lightweight cryptographic protocols.\\
\noindent\textbf{Strategic Implications.}  
Rather than supplant existing post-quantum cryptographic primitives (e.g., lattice-based schemes, ECC, or RSA), the proposed approach serves as a modular entropy conditioner and validation filter. It ensures that critical seed material -- such as secret keys, IVs, or nonces -- meets algebraic admissibility constraints while maintaining high entropy. In doing so, it strengthens the foundational layer of cryptographic systems and enhances their robustness against both statistical flaws and physical leakage channels.\\
Further extensions of this framework, including its generalisation to broader algebraic settings and integration into post-quantum cryptographic stacks, are currently under investigation.\\
In conclusion, the cyclic modular inversion strategy presented here offers a deterministic, entropy-aware, and algebraically sound foundation for cryptographic seed generation. It strengthens the design of secure systems by marrying symbolic number theory with practical cryptographic demands -- particularly in embedded, low-power, or post-quantum environments.

\section*{License}
This manuscript is licensed under a Creative Commons Attribution-NonCommercial-NoDerivatives 4.0 International License (CC BY-NC-ND 4.0). The author retains all rights, including patent rights.

\section*{Acknowledgements}
The author acknowledges that elements of this work are covered under UK Patent Application No.~GB2508035.9, titled ``System and Method for Modular Residue-Based Seed Filtering and Entropy Generation in Cryptographic Applications.'' The application was filed in 2025 and is currently pending; it has not yet been made publicly available.

\appendix
\section{Appendix A: Glossary of Terms and Acronyms}
\begin{table}[H]
\centering
\caption{Glossary of technical terms and symbols used in the paper.}
\begin{tabular}{|p{3cm}|p{10cm}|}
\hline
\textbf{Term / Symbol} & \textbf{Description} \\
\hline
$3^p$ & Prime power modulus used for residue computation. \\
$\mathbb{Z}/3^p\mathbb{Z}$ & Ring of integers modulo $3^p$. \\
$\mathbb{Z}/3^p\mathbb{Z}^*$ & Multiplicative group of units modulo $3^p$. \\
$k$ & Integer index representing exponent in $2^{k-1}$. \\
ECS & Entropy Confidence Score – a composite metric for evaluating residue randomness and balance. \\
CD & Cycle Density – proportion of the multiplicative group covered by the sequence. \\
RUD & Residue Uniformity Deviation – quantifies deviation in frequency distribution of residues. \\
MBI & Modular Bias Index – measures statistical skew across residues modulo $3^p$. \\
DRBG & Deterministic Random Bit Generator. \\
PRNG & Pseudorandom Number Generator. \\
IND-CPA & Indistinguishability under Chosen Plaintext Attack. \\
\hline
\end{tabular}
\end{table}

\bibliographystyle{unsrt}
\bibliography{cryptoseed_arxiv_idowu}

\begin{thebibliography}{10}

\bibitem{Knuth1997}
Donald~E. Knuth.
\newblock The art of computer programming, volume 2: Seminumerical algorithms.
\newblock {\em Addison-Wesley}, 1997.
\newblock Book.

\bibitem{Kocher1999}
Paul Kocher, Joshua Jaffe, and Benjamin Jun.
\newblock Differential power analysis.
\newblock {\em CRYPTO}, 1999.
\newblock Conference paper.

\bibitem{rivest1978}
Ronald~L. Rivest, Adi Shamir, and Leonard Adleman.
\newblock A method for obtaining digital signatures and public-key
  cryptosystems.
\newblock {\em Communications of the ACM}, 21(2):120--126, 1978.

\bibitem{miller1986}
Victor~S. Miller.
\newblock Use of elliptic curves in cryptography.
\newblock {\em Advances in Cryptology--CRYPTO '85}, pages 417--426, 1986.
\newblock Conference paper.

\bibitem{Szabo1967}
Nicholas~S. Szabo and Richard~I. Tanaka.
\newblock Residue arithmetic and its applications to computer technology.
\newblock {\em McGraw-Hill}, 1967.
\newblock Book.

\bibitem{Lyubashevsky2010}
Vadim Lyubashevsky, Chris Peikert, and Oded Regev.
\newblock On ideal lattices and learning with errors over rings.
\newblock {\em EUROCRYPT}, 2010.
\newblock Conference paper.

\bibitem{Idowu2025}
Michael~A. Idowu.
\newblock Symbolic generation and modular embedding of high-quality
  abc-triples, 2025.

\bibitem{koblitz1994}
Neal Koblitz.
\newblock A course in number theory and cryptography.
\newblock {\em Springer}, 1994.
\newblock Book.

\bibitem{Bernstein2017}
Daniel~J. Bernstein and Tanja Lange.
\newblock Post-quantum cryptography.
\newblock {\em Nature}, 549(7671):188--194, 2017.

\bibitem{Gallian2021}
Joseph~A. Gallian.
\newblock Contemporary abstract algebra.
\newblock {\em Cengage Learning}, 2021.
\newblock 10th edition, Book.

\bibitem{ireland1990}
Kenneth Ireland and Michael Rosen.
\newblock A classical introduction to modern number theory.
\newblock {\em Springer}, 1990.
\newblock 2nd edition, Book.

\bibitem{Kelsey2015}
John Kelsey.
\newblock Entropy and security in cryptographic systems.
\newblock Presented at the Real-World Cryptography Workshop (RWC), 2015.
\newblock Available at: \url{https://rwc.iacr.org/2015/Slides/kelsey.pdf}.

\bibitem{Nist90}
John Kelsey, Lawrence Bassham, Meltem~S. Turan, and Elaine Barker.
\newblock Recommendation for the entropy sources used for random bit
  generation.
\newblock Technical Report NIST Special Publication 800-90B, National Institute
  of Standards and Technology (NIST), Gaithersburg, MD, 2018.

\end{thebibliography}

\end{document}